\RequirePackage{fix-cm}
\documentclass[smallextended]{svjour3}       
\smartqed  
\usepackage{float}
\usepackage[authoryear]{natbib}
\usepackage{graphicx}
\usepackage{hyperref}
\usepackage{xcolor}
\usepackage[most]{tcolorbox}
\usepackage{microtype}     
\usepackage{multirow}
\usepackage{makecell}
\usepackage{caption}
\usepackage{tabularx}
\usepackage{booktabs}
\usepackage{xcolor}
\usepackage{xurl}
\colorlet{diffred}{red!80!black}
\colorlet{diffgreen}{green!60!black}

\newcommand{\diffdel}[1]{\textcolor{diffred}{\texttt{- #1}}}
\newcommand{\diffadd}[1]{\textcolor{diffgreen}{\texttt{+ #1}}}
\newcommand{\diffctx}[1]{\textcolor{gray}{\texttt{\ \ #1}}} 
\usepackage{tikz}
\usetikzlibrary{arrows.meta,calc}
\usepackage{enumitem}
\usepackage{tikz}
\usetikzlibrary{calc,arrows.meta}
\usepackage{graphicx} 
\usepackage{tikz}
\usetikzlibrary{positioning}

\usepackage{booktabs,tabularx}

\definecolor{darkgreen}{rgb}{0.0, 0.5, 0.0}

\newcommand{\rqone}{\textit{What are the characteristics of cache-adopting repositories?}}
\newcommand{\rqtwo}{\textit{How do caching configurations evolve over time across different CI/CD job types? }}
\newcommand{\rqthree}{\textit{Why do developers make cache-related modifications in GitHub Actions workflows?}}

\begin{document}

\title{How Developers Adopt, Use, and Evolve CI/CD Caching: An Empirical Study on GitHub Actions} 

\titlerunning{How Developers Adopt, Use, and Evolve CI/CD Caching}

\author{
        Kazi Amit Hasan \and
        Yuan Tian \and
        Safwat Hassan \and
        Steven H. H. Ding 
}




\institute{
  Kazi Amit Hasan, Yuan Tian \at
  School of Computing, Queen's University, ON, Canada \\
  \email{\{kaziamit.hasan, y.tian\}@queensu.ca}
  \and
  Safwat Hassan \at
  Faculty of Information, University of Toronto, ON, Canada \\
  \email{safwat.hassan@utoronto.ca}
  \and
  Steven H. H. Ding \at
  School of Information Studies, McGill University, Canada \\
  \email{steven.h.ding@mcgill.ca}
}

\date{Received: date / Accepted: date}

\maketitle

\begin{abstract}


Continuous Integration/Continuous Delivery (CI/CD) caching is widely used to reduce repeated computation and improve CI/CD efficiency, yet maintaining effective caching requires ongoing maintenance effort. In this paper, we present the first empirical study on how developers configure and evolve caching in CI/CD workflows on GitHub Actions. We analyze 952 GitHub repositories (266 cache adopters and 686 non-adopters), to compare repository characteristics, characterize caching usage at the job and step levels, uncover patterns in caching configuration evolution, and identify the drivers of cache-related changes. Our analysis spans 1,556 workflow files, 10,373 commits, and 17,185 workflow configuration changes, including an average of 9.37 cache-related changes per repository. Our main observations are: (1) cache-adopting repositories are more active and popular than non-adopters; (2) caching is used across multiple CI/CD job types through a variety of caching mechanisms rather than a single standardized approach; (3) caching configurations evolve through frequent, repetitive maintenance patterns, with rapid updates in build and test jobs and slower evolution in other job types; and (4) cache-related modifications are driven by distinct maintenance needs: parameter updates are mainly human-driven to fix issues, while version updates occur later and are often bot-driven for dependency maintenance. Our findings quantify the substantial maintenance effort involved in CI/CD caching and highlight opportunities to improve reliability and tool support.


\end{abstract}

\maketitle

\section{Introduction} \label{sec:intro}

Modern software development relies on Continuous Integration and Continuous Deployment (CI/CD) pipelines to automate integration, testing, and deployment activities, enabling teams to sustain fast iteration and gradually improve product quality~\citep{shahin2017continuous, rahman2023exploring, yang2025impact}. A variety of platforms support CI/CD, including Jenkins, Travis CI, CircleCI, and GitHub Actions (GHA)~\citep{ghaleb2026ci}. As projects grow, CI/CD workflows scale in complexity, increasing in the number of jobs (sets of steps executed on dedicated runners), the volume of dependencies, and execution frequency. This growth makes repeated work, such as reinstalling dependencies and rebuilding intermediate outputs, a common source of slowdown.\footnote{https://runs-on.com/github-actions/caching-dependencies/} Even small inefficiencies accumulate into substantial delays and computational overhead when runs are frequent, extending feedback cycles and increasing computational costs ~\citep{gallaba2020accelerating,bouzenia2024resource}. To address this, CI/CD platforms, such as GHA provide configurable caching mechanisms that allow developers to store and reuse artifacts ~\citep{gallaba2020accelerating, ghaleb2026promise}. For many projects, caching is not just a performance optimization but a necessary mechanism to maintain fast development cycles as CI/CD workflows grow in size, complexity, and frequency~\citep{gallaba2019improving}. 

However, CI/CD caching is not a simple toggle. Implementing an effective caching strategy is often complex because developers must determine when caching should be applied, what artifacts should be stored, and how cached artifacts should be identified and reused across workflow runs. These decisions are critical for ensuring that cached artifacts are reused when appropriate while avoiding the reuse of outdated artifacts. For example, an improperly designed key can cause repeated cache misses\footnote{https://www.tencentcloud.com/techpedia/130942} and provide little performance benefit\footnote{https://aws.amazon.com/caching/best-practices/}. As a result, what is intended as a performance boost\footnote{https://blog.jetbrains.com/teamcity/2025/12/is-your-ci-cd-tool-helping-or-hindering-performance/} often becomes a non-trivial maintenance burden that adds to the hidden costs of CI/CD~\citep{valenzuela2024hidden}. This burden is evident in real workflow maintenance activities, such as adjusting cache keys and paths, adding or removing caches when they prove ineffective, and updating cache-related configurations to remain compatible over time. Despite this practical complexity, caching has received limited attention in prior literature work as a configuration and maintenance concern in CI/CD workflows. 


In this work, we aim to better understand CI/CD caching, particularly the maintenance of caching configurations, by focusing on GHA, given its popularity and rich caching mechanisms. Prior studies have examined GHA workflows and their evolution~\citep{chen2021let,decan2022use}, but provide limited fine-grained analysis of caching itself. They treat caching as a simple CI/CD optimization, focusing primarily on whether it is enabled and reporting it as one of the most commonly used strategies. As a result, caching is often viewed as a one-time configuration choice rather than an ongoing maintenance concern, leaving its configuration and maintenance dynamics poorly understood. This gap is practically significant, as practitioners may underestimate the effort required to keep caching effective as projects evolve.

To fill this gap, we present the first large-scale, cache-centric empirical study of GHA workflows. We analyze caching configurations and their evolution in 266 cache-adopting repositories, encompassing 1,556 workflow files, 10,373 commits, and 17,185 workflow configuration changes (2494 of them are caching-related), drawn from a broader sample of 952 GitHub repositories that use GHA. Our study addresses three research questions: 

\begin{itemize}
\item [] \textbf{RQ1: \rqone} We find that cache-adopting repositories are more active and popular than non-adopters. Caching is used across diverse CI/CD job types, with build and test jobs being the most prominent. At the step level, developers employ multiple caching mechanisms, where explicit caching via \texttt{actions/cache} (70.9\%) dominates, followed by package manager (24.5\%) and Docker layer caching (0.2\%).

\vspace{0.1cm}
\item [] \textbf{RQ2: \rqtwo} We find that caching evolution exhibits both shared patterns and job-specific differences. Across job types, caching configurations evolve through frequent, iterative updates and corrective actions, such as removing and immediately re-adding caches. Build and test jobs exhibit rapid, repeated parameter updates, whereas job types such as release exhibit slower modifications to caching configurations, with longer gaps between changes.

\vspace{0.1cm}
\item [] \textbf{RQ3: \rqthree} By linking observed quantitative caching modification patterns with qualitative evidence from commit messages and pull requests, we develop a taxonomy of the drivers behind these changes. We find that parameter updates are primarily human-driven and aimed at fixing caching issues, whereas cache version updates are often triggered by bot alerts and repository dependency maintenance activities. In general, developers perform most of the debugging, tuning, and restructuring that shape how caching evolves in practice.


\end{itemize}
\noindent \textbf{Our contributions:}
\begin{itemize}
\item We present the first large-scale empirical study of CI/CD caching in GHA, characterizing its adoption across repositories and languages, the evolution of caching configurations over time, and the drivers behind these changes. 

\item We broaden prior work, which primarily focuses on explicit caching via \texttt{actions/cache}, by capturing implicit caching mechanisms such as package manager and Docker layer caching, providing a more comprehensive view of how caching is implemented in GHA workflows.

\item We quantify how caching configurations evolve over time by representing cache-related maintenance activities as state transitions and analyzing their evolution patterns across different job types.

\item We explain why cache-related modifications occur by linking observed transitions to commit messages and PR context and constructing a taxonomy of drivers grounded in that evidence.

\item To foster future research in the area, we have made our replication package publicly available on GitHub.\footnote{https://github.com/RISElabQueens/caching-in-github-actions}
\end{itemize}
We organize the remainder of the paper as follows. Section~\ref{sec:dataset} introduces our dataset. Sections~\ref{sec:rq1} to~\ref{sec:rq3} present the methodology and answers to each research question. Section~\ref{sec:disc} discusses implications for practitioners and researchers. Section~\ref{sec:threats} presents threats to validity, and Section~\ref{sec:related} presents related work. Finally, Section~\ref{sec:conclusion} concludes the paper.

\section{Background} \label{sec:background}
\subsection{Caching in GitHub Actions}
GitHub Actions defines CI and CD processes through workflow files written in YAML and stored under the \texttt{.github/workflows} directory of a repository. A workflow can contain one or more jobs. Each job runs on a runner that provides the execution environment, and each job consists of an ordered sequence of steps that perform tasks such as checking out code, setting up languages, installing dependencies, building, and testing.\footnote{https://docs.github.com/en/actions/get-started/understand-github-actions} Caching is a performance optimization mechanism that reuses files across workflow runs to reduce repeated work, such as reinstalling dependencies and rebuilding intermediate artifacts. In GHA, caching is typically configured as a dedicated step within a job using the available caching strategies. Cache configuration requires two user-specified parameters,  \texttt{path} and \texttt{key}. The \texttt{path} identifies the file(s),  directory(ies) to be cached, while the \texttt{key} serves as a unique identifier for locating and restoring cached artifacts across runs. 

\begin{figure*}[h]
    \centering
    \includegraphics[width=1.0\linewidth]{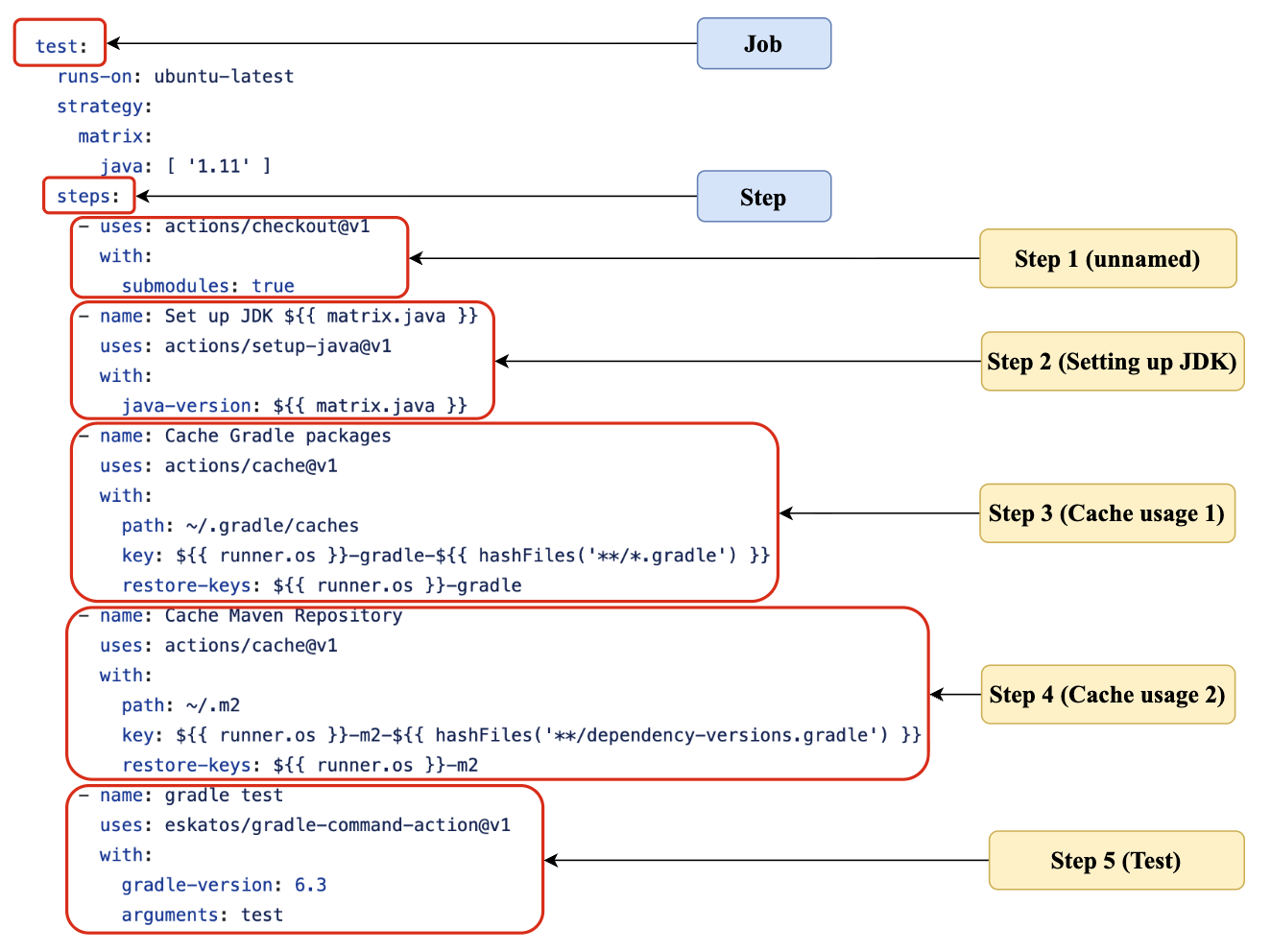}
    \caption{An example of GitHub Actions workflow illustrating jobs, steps, and cache steps.}
    \label{fig:example}
\end{figure*}

Figure \ref{fig:example} shows an example workflow that illustrates how jobs and steps are structured and how caching appears as steps inside a job\footnote{\url{https://github.com/apache/incubator-tuweni/blob/8bb12d442f6ac43e1ea616e0cc2e5d1fcb1e73ee/.github/workflows/master-pr-build.yml}}. The workflow defines a single job named test that runs on ubuntu-latest and uses a Java version from a small matrix. The job first checks out the repository, then sets up the JDK using \texttt{actions/setup-java}. It then includes two cache steps using \texttt{actions/cache}.One cache step stores Gradle cache files in the \texttt{\textasciitilde/.gradle/caches} directory, while the other stores the local Maven repository in the \texttt{\textasciitilde/.m2} directory. Both cache steps define cache keys that include the runner operating system and a hash of dependency-related files. This setup illustrates how a job can include multiple caches and how keys can be tied to dependency definitions so that caches remain stable when dependencies do not change and are refreshed when dependency files are modified.

\subsection{Cache-related maintenance activities and terminology}

To support our empirical analysis, we define terminology related to caching in GHA workflows. At the repository level, we distinguish between \textit{cache-adopting} and \textit{non-adopters} repositories. A repository is considered cache-adopting if it enables caching at any point during its lifecycle. Conversely, a repository is considered a non-adopter (for caching) if none of its workflows contain any cache configuration throughout the observed history. This repository-level classification is used in RQ1 to compare adopters and non-adopters, while RQ2 and RQ3 analyze cache-related changes only within repositories that adopt caching.  Next, we define \textit{cache-related maintenance activities} to capture how developers introduce, update, and manage caching in GHA workflows (Table~\ref{tab:cache_maintenance}). These activities serve as a common vocabulary across the paper and provide a consistent basis for repository characterization (RQ1), evolution analysis (RQ2), and driver analysis (RQ3). 


\newcolumntype{Y}{>{\raggedright\arraybackslash}X}

\begin{table}[H]
\caption{Taxonomy of cache-related maintenance activities.}
\label{tab:cache_maintenance}
\centering
\scriptsize 
\setlength{\tabcolsep}{2pt} 
\renewcommand{\arraystretch}{1.1}

\begin{tabularx}{\linewidth}{@{} l l l >{\hsize=0.85\hsize}Y >{\hsize=1.15\hsize}Y @{}}
\toprule
\textbf{Category} & \textbf{Abbr.} & \textbf{Sub-category} & \textbf{Definition} & \textbf{Example (Diff)} \\
\midrule

\begin{tabular}[t]{@{}l@{}}Cache\\enablement\end{tabular}
& EC
&
& Introduction of caching for the first time in a workflow, corresponding to the initial addition of a cache step within a job.
& \parbox[t]{\linewidth}{\raggedright
\diffdel{(no cache step)}\\
\diffadd{uses: actions/cache@v3}\\
\diffctx{with:}\\ 
\diffctx{\ \ path: \~/.cache/pip}
} \\
\midrule

\begin{tabular}[t]{@{}l@{}}Cache\\version\\update\end{tabular}
& C\_up
&
& Any modification to the cache action version specification.
& \\
\cmidrule(lr){3-5}
& & Update
& Refers to modifications within the caching version configuration.
& \parbox[t]{\linewidth}{\raggedright
\diffdel{uses: actions/cache@v3}\\
\diffadd{uses: actions/cache@wgewetwetwe}
} \\
\cmidrule(lr){3-5}
& & Upgrade
& Represents incremental cache version updates, where developers move to a newer version of cache.
& \parbox[t]{\linewidth}{\raggedright
\diffdel{uses: actions/cache@v2}\\
\diffadd{uses: actions/cache@v3}
} \\
\cmidrule(lr){3-5}
& & Downgrade
& Occurs when developers revert to an older version within caching configurations.
& \parbox[t]{\linewidth}{\raggedright
\diffdel{uses: actions/cache@v3}\\
\diffadd{uses: actions/cache@v2}
} \\
\midrule

\begin{tabular}[t]{@{}l@{}}Adding a\\new cache\end{tabular}
& C\_add
&
& Introduction of an additional cache step alongside existing caches, typically to support other jobs or workflow stages.
& \parbox[t]{\linewidth}{\raggedright
\diffdel{(cache already exists)}\\
\diffadd{(adding additional cache)}\\
\diffadd{uses: actions/cache@v3}\\
\diffctx{with:}\\
\diffctx{\ \ path: node\_modules}
} \\
\midrule

\begin{tabular}[t]{@{}l@{}}Cache\\removal\end{tabular}
& C\_rm
&
& Removal or disabling of a previously defined cache step when it is no longer beneficial.
& \parbox[t]{\linewidth}{\raggedright
\diffdel{uses: actions/cache@v3}\\
\diffdel{with:}\\
\diffdel{\ \ path: node\_modules}\\
\diffadd{(cache step removed)}
} \\
\midrule

\begin{tabular}[t]{@{}l@{}}Parameter\\update\end{tabular}
& P\_up
&
& Modification of existing cache parameter values.
& \\
\cmidrule(lr){3-5}
& & Update
& Refers to textual modifications within the caching configuration parameters.
& \parbox[t]{\linewidth}{\raggedright
\diffdel{key: \${{ runner.os }}-deps-\${{ hashFiles('**/yarn.lock') }}}\\
\diffadd{key: \${{ runner.os }}-deps-\${{ hashFiles \\ ('**/package-lock.json') }}}
} \\
\cmidrule(lr){3-5}
& & Upgrade
& Refers incremental language version updates.
& \parbox[t]{\linewidth}{\raggedright
\diffdel{with: \{ node-version: 16 \}}\\
\diffadd{with: \{ node-version: 18 \}}
} \\
\cmidrule(lr){3-5}
& & Downgrade
& Refers to parameter specific version downgrades.
& \parbox[t]{\linewidth}{\raggedright
\diffdel{with: \{ python-version: '3.12' \}}\\
\diffadd{with: \{ python-version: '3.11' \}}
} \\
\midrule

\begin{tabular}[t]{@{}l@{}}Parameter\\addition\end{tabular}
& P\_add
&
& Introduction of a new cache parameter that was not previously specified.
& \parbox[t]{\linewidth}{\raggedright
\diffdel{(no parameters)}\\
\diffadd{restore-keys: \${{ runner.os }}-deps-}
} \\
\midrule

\begin{tabular}[t]{@{}l@{}}Parameter\\removal\end{tabular}
& P\_rm
&
& Removal of an existing cache parameter that is no longer required.
& \parbox[t]{\linewidth}{\raggedright
\diffdel{restore-keys: \${{ runner.os }}-deps-}\\
\diffadd{(restore-keys removed)}
} \\
\bottomrule
\end{tabularx}
\end{table}

\section{Data Collection and Preparation}\label{sec:dataset}

\begin{figure*}[]
    \centering
    \includegraphics[width=1\linewidth]{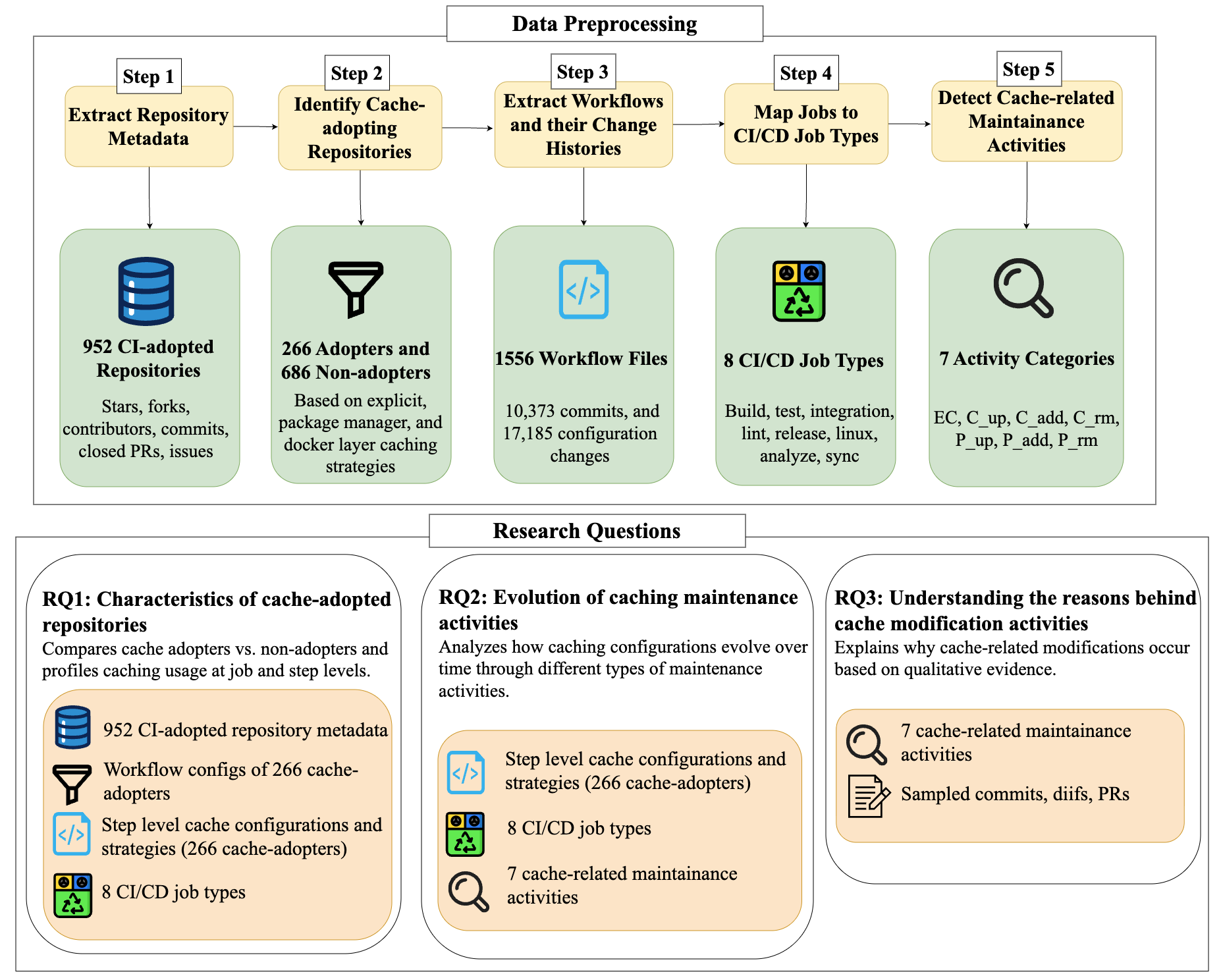}
    \caption{ Overview of our study.}
    \label{fig:overall}
\end{figure*}

In this section, we describe how we collected and prepared data for our empirical study. Figure \ref{fig:overall} presents an overview of the study design and illustrates how each step of the process and the resulting data artifacts contribute to each research question. 

\begin{itemize}
    \item [] \noindent \textbf{Step 1: Extract repository metadata:}
Our study builds on the dataset curated by Bouzenia et al.~\citep{bouzenia2024resource}, which contains 952 GitHub repositories that use GitHub Actions (GHA) as their CI/CD automation platform. For each repository, we collect repository-level metadata through the GitHub GraphQL API.\footnote{https://docs.github.com/en/graphql} The collected metadata includes the numbers of stars, forks, contributors, commits, closed pull requests, and issues, together with the repository's primary programming language. This step produces the repository metadata artifact shown in Figure~\ref{fig:overall}, which is later used in RQ1 to compare cache adopters and non-adopters.

\vspace{0.2cm}

\item [] \noindent \textbf{Step 2: Identify cache adopting repositories:} Our analysis focuses on caching configurations in GHA workflows. We label a repository as cache-adopting if caching is enabled in any job at any point in its workflow history. We refer to the first such event as cache enablement (EC), following the taxonomy in Table~\ref{tab:cache_maintenance}. We capture caching through three cases\footnote{https://www.ubicloud.com/blog/github-actions-transparent-cache}. First, we capture explicit caching, where a job includes an actions/cache step and developers specify its inputs (e.g., path and key). Second, we capture package manager caching, where caching is enabled through language/tool setup actions (e.g., setup-go, setup-node, setup-python) using their caching settings. Third, we capture docker layer caching, where caching is configured through Docker build tooling (e.g., docker/build-push-action) to reuse previously built layers. As caching behavior differs across setup actions and versions, we follow official documentation\footnote{https://github.com/actions/cache} and treat caching as present only when it is actually enabled in the workflow configuration. In particular, actions/setup-go@v4 enables caching by default, while earlier versions of setup-go do not. Therefore, for Go, we treat caching as enabled by default only when the workflow uses actions/setup-go@v4 and anything after this. For other setup actions in our scope, caching is not enabled by default in the versions we observe, so we require an explicit caching configuration in the action inputs (e.g., enabling the caching option). Table~\ref{tab:caching_default} summarizes the setup actions considered and whether caching is enabled by default in the action version we observe in workflow files.

\begin{table}[]
\centering
\caption{Cache-related actions considered in our study and their default configurations.}
\label{tab:caching_default}
\scriptsize
\begin{tabular}{llll}
\toprule
\textbf{Caching Strategy} & \textbf{Ecosystem} & \textbf{Actions} & \textbf{\shortstack{Default Caching}} \\
\midrule
\multirow{1}{*}{Explicit caching} 
& General & \texttt{actions/cache} & No \\

\midrule
\multirow{5}{*}{Package manager caching} 
& Go & \texttt{actions/setup-go@v4} & Yes\textsuperscript{1} \\
& Python & \texttt{actions/setup-python} & No \\
& Node.js & \texttt{actions/setup-node} & No \\
& Ruby & \texttt{ruby/setup-ruby} & No \\
& Java & \texttt{actions/setup-java} & No \\

\midrule
\multirow{1}{*}{Docker layer caching}
& Docker & \texttt{docker/build-push-action} & No \\

\bottomrule
\end{tabular}


\vspace{0.5cm}
{\footnotesize\raggedright
\textsuperscript{1}Caching is enabled by default only for \texttt{actions/setup-go@v4}, earlier versions do not enable caching by default.\par
}
\end{table}

Applying this identification procedure to the 952 repositories yields 266 cache-adopted repositories (27.9\% of the dataset) that enabled caching at any point during their lifecycle, as recorded in their commit history, and 686 non-adopters.We retain both groups for RQ1. The 266 cache adopters then proceed to the subsequent preprocessing steps, because only these repositories contain caching configurations that can be examined longitudinally.

\vspace{0.2cm}

\item [] \noindent \textbf{Step 3: Extract workflows and their change histories.}
For the 266 cache adopters, we reconstruct workflow histories from Git history using Gigawork\footnote{\url{https://github.com/cardoeng/gigawork}}. Gigawork extracts historical versions of workflow files together with commit-level metadata, including commit hash, timestamp, and file path. Our reconstructed workflow history includes commits up to May 5, 2025. This step yields 1,556 workflow files, 10,373 commits, and 17,185 configuration changes, as shown in Figure~\ref{fig:overall}. These reconstructed histories allow us to trace how caching-related configurations evolve over time. Although the source dataset also contains workflow-run logs, our analyses focus on repository metadata and workflow configuration histories, as these are the data required to answer our research questions.

\vspace{0.2cm}

\item [] \noindent \textbf{Step 4: Map jobs to CI/CD job types.}
To support cross-repository analysis, we normalize developer-defined job names into eight CI/CD job types, adapting the taxonomy of \citet{bouzenia2024resource}. The resulting categories are \textit{build}, \textit{test}, \textit{integration}, \textit{lint}, \textit{release}, \textit{linux}, \textit{analyze}, and \textit{sync}. This normalization is necessary because GHA job names are not standardized across repositories, and direct aggregation of raw job names would not support meaningful cross-project comparison.

\vspace{0.2cm}

\item [] \noindent \textbf{Step 5: Detect cache-related maintenance activities.}
Finally, we analyze workflow histories and retain only changes that modify caching-related configurations. Through this process, we identify 2,494 cache-related changes, representing 14.53\% of all workflow changes.  Each such changes are mapped to one of the seven cache-related maintenance activity categories defined in Table~\ref{tab:cache_maintenance}: \textit{EC}, \textit{C\_up}, \textit{C\_add}, \textit{C\_rm}, \textit{P\_up}, \textit{P\_add}, and \textit{P\_rm}. Workflow changes unrelated to caching are excluded at this stage. This step produces the activity-level dataset used to model cache evolution and to support the qualitative analysis of cache-related modifications.

\vspace{0.2cm}
\end{itemize}

\noindent \textbf{Preparing data for RQ1:}
RQ1 examines the characteristics of cache adopters and profiles how caching is used within adopting repositories. As shown in Figure~\ref{fig:overall}, RQ1 combines four artifacts: (1) repository metadata for all 952 repositories, (2) workflow configurations of the 266 cache adopters, (3) step-level cache configurations and strategies extracted from those adopters, and (4) the mapping of jobs to the eight CI/CD job types. Repository metadata is used to compare adopters and non-adopters, while workflow and step-level configurations are used to characterize caching prevalence and caching strategy usage at both the job and step levels. 

\vspace{0.2cm}
\noindent \textbf{Preparing data for RQ2:}
RQ2 investigates how caching configurations evolve through different maintenance activities over time. For this analysis, we use the step-level cache configurations of the 266 cache adopters, the eight CI/CD job types, and the seven cache-related maintenance activity categories. After filtering workflow histories to retain only cache-related changes, we reconstruct per-job sequences of cache maintenance activities over time and group them by the CI/CD job types.

\vspace{0.2cm}

\noindent \textbf{Preparing data for RQ3:}
RQ3 explains why cache-related modifications occur. This analysis is built on the cache-related commits identified in Step 5. We group them by their target activity type \textit{(e.g., C\_up, P\_up, C\_add, P\_add, C\_rm, P\_rm)} and select a statistically significant random
sample from each group (sample sizes reported in the RQ3 section). For each sampled commit, we collect the workflow-file diff, the commit message, and the linked pull request when available. We also record whether the change was introduced by a human or a bot based on author identity and automation indicators. These artifacts provide the qualitative evidence used to interpret the reasons behind cache-related modification activities.

\section{RQ1: \rqone}\label{sec:rq1}
\subsection{Motivation}


This research question aims to understand how caching is configured in GHA workflows and how cache-adopting repositories differ from non-adopters. Answering this question provides insight into when caching becomes necessary in practice and the common configuration patterns that developers adopt. 


\subsection{Approach}
To characterize repositories that adopt GHA caching, we combine a repository-level comparison with a workflow-level analysis. Specifically, at the repository level, we compare cache-adopting and non-adopters using metadata (e.g., contributors, commits, stars) collected via the GitHub GraphQL API (ref. Section~\ref{sec:dataset}). For cache-adopting repositories, we further analyze their CI/CD workflows using the latest repository snapshots to capture their current configurations. We quantify workflow complexity by summarizing the number of workflow files and jobs per repository.

To characterize caching practices, we conduct an analysis of caching configurations at two levels of granularity:
\begin{itemize}
\item[] \textbf{Job-level:} We categorize the jobs specified in GHA workflow files into high-level types based on their developer-defined names using the keyword-based taxonomy proposed by Bouzenia et al.~\citep{bouzenia2024resource}. For each job, we determine whether caching is used by checking for the presence of at least one cache-related configuration step. We then compute caching prevalence for each job type as the percentage of jobs that contain caching.

\vspace{0.2cm}

\item[] \textbf{Step-level:} We extract cache-related configuration steps from workflow files and categorize them based on the caching mechanism used (e.g., explicit caching, package manager caching, and docker layer caching). We further analyze their distribution across job types. 

\end{itemize}

\subsection{Results} The following observations summarize patterns identified from our analysis of the 266 cache-adopted repositories (1,556 YAML files) and 686 non-adopting repositories in our dataset. 

\begin{figure*}
    \centering
    \includegraphics[width=1.0\linewidth]{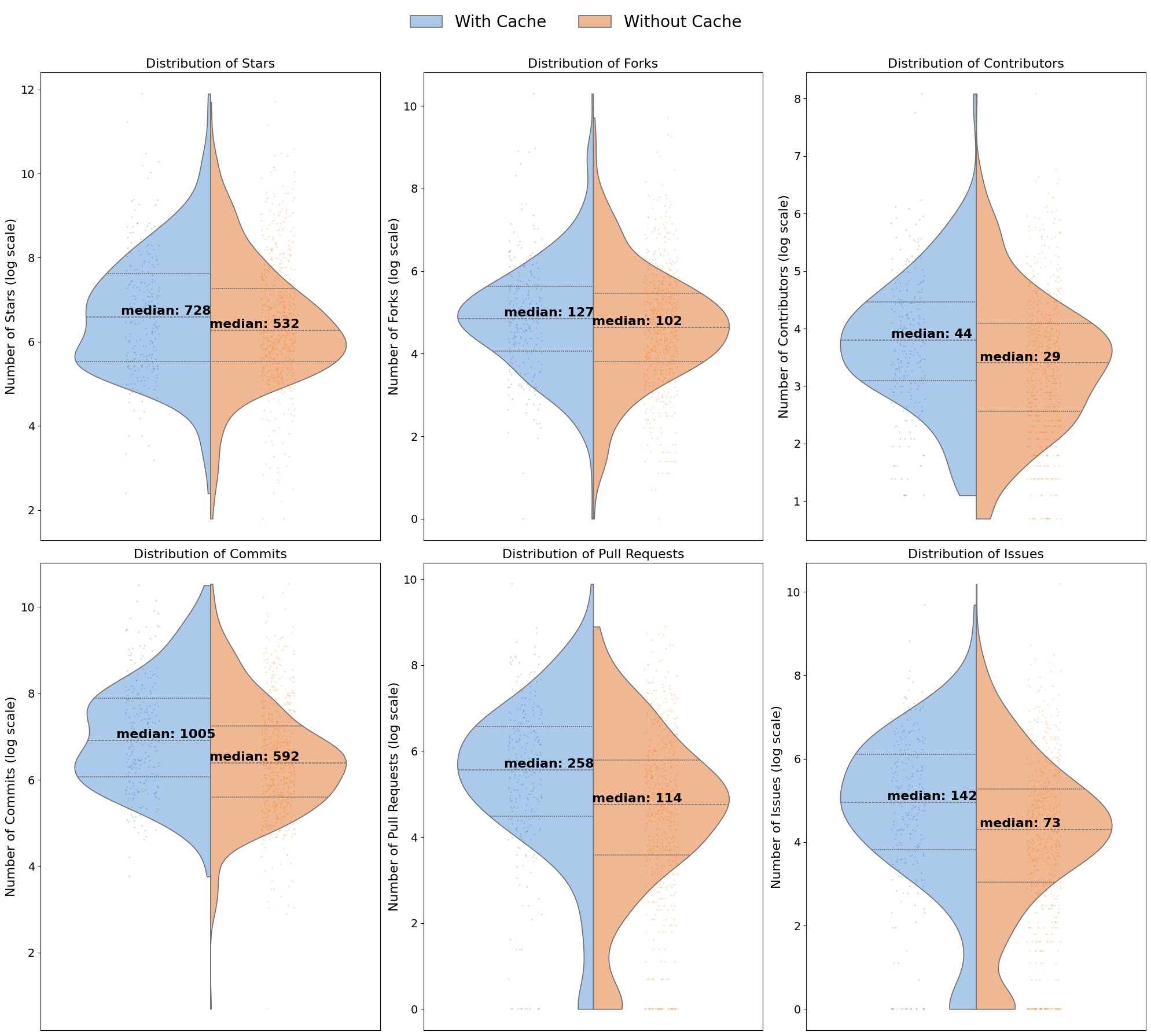}
   \caption{Distributions of repository characteristics for cache-adopting (with) and non-adopting (without) repositories. Values are plotted on a log scale, with medians annotated in the original scale.}
    \label{fig:doublesided}
\end{figure*}

\vspace{0.1cm}
\noindent \textbf{Observation 1.1: \textit{Cache-adopting repositories tend to be more active, collaborative, and popular than non-adopting repositories.}} 
Figure \ref{fig:doublesided} shows the repository-level characteristics of cache-adopting and non-adopting repositories. Compared with non-adopters, cache-adopting repositories have more contributors, commits, closed PRs, stars, forks, and issues. They also exhibit greater community interest, as indicated by higher median numbers of stars and forks. These results suggest that repositories with higher activity and community engagement are more likely to adopt caching to help manage CI workloads.

\vspace{0.2cm}

\noindent \textbf{Observation 1.2: \textit{Caching adoption differs across primary languages.}} Table~\ref{tab:language_stats} compares the primary programming languages of cache-adopting repositories and non-adopters in our dataset. To account for differences in language prevalence in our sample, we report the \emph{within-language adoption rate}, which captures the proportion of repositories in each language category that adopt caching. PHP and Java show the highest adoption rates (54.2\% and 50.7\%), while Python (22.3\%), Ruby (17.6\%) show comparatively lower adoption. 

A chi-square test of independence shows that caching adoption is significantly associated with programming language ($p < 0.001$). This result indicates that adoption rates of caching differ significantly across languages rather than being uniformly distributed.


\begin{table}[]
\centering
\caption{Adoption rates of caching across primary programming languages.}
\label{tab:language_stats}
\begin{tabular}{lrrr}
\toprule
\textbf{Language} & \textbf{Cache-adopting} & \textbf{Non-adopters} & \textbf{Adoption Rate} \\ 
\midrule
PHP & 26 & 22 & 54.2\% \\ 
Java & 36 & 35 & 50.7\% \\ 
Kotlin & 12 & 21 & 36.4\% \\ 
TypeScript & 45 & 79 & 36.3\% \\ 
Go & 22 & 54 & 28.9\% \\ 
JavaScript & 37 & 109 & 25.3\% \\ 
C++ & 17 & 52 & 24.6\% \\ 
Python & 42 & 146 & 22.3\% \\ 
C & 9 & 38 & 19.1\% \\ 
Shell & 4 & 17 & 19.0\% \\ 
Ruby & 6 & 28 & 17.6\% \\ 
Others & 10 & 79 & 11.2\% \\ 
\bottomrule
\end{tabular}
\end{table}
\vspace{0.2cm}

\noindent \textbf{Observation 1.3: \textit{Cache-adopting repositories typically maintain a moderate number of workflow files and jobs, with substantial long-tail complexity.}} Table \ref{tab:dist_files_jobs} presents the per-repository workflow structure of cache-adopting repositories. The median number of workflow files per repository is 3, indicating that most adopters maintain multiple workflow files rather than a single configuration. While the minimum is 1 file, the maximum reaches 75 files, and the standard deviation (8.02) exceeds the mean (5.85), suggesting a right-skewed distribution with a long tail of repositories exhibiting substantially more complex workflow setups.

A similar pattern is observed in jobs. The median number of jobs per repository is 4, with an average of 6.57 jobs. Although many repositories operate with a relatively small number of jobs, the maximum reaches 116, indicating the presence of a small subset of repositories with highly complex CI/CD configurations.


\begin{table}[h]
\centering
\caption{Distribution of workflow files and jobs per repository in cache-adopting projects.}
\label{tab:dist_files_jobs}
\resizebox{\linewidth}{!}{%
\begin{tabular}{lrrrrrrr}
\toprule
\textbf{Metric} & \textbf{Min} & \textbf{Mean} & \textbf{Std} & \textbf{25\%} & \textbf{50\%} & \textbf{75\%} & \textbf{Max} \\ 
\midrule
\# of files per repository & 1 & 5.85 & 8.02 & 2 & 3 & 6 & 75 \\ 
\# of jobs per repository & 1 & 6.57 & 11.14 & 2 & 4 & 7 & 116 \\ 
\bottomrule
\end{tabular}%
}
\end{table}

\vspace{0.2cm}

\noindent \textbf{Observation 1.4: \textit{Caching is most prevalent in build, integration, lint, and test jobs.}} Table \ref{tab:cicd_phase} reports the highest job-level caching prevalence in build (37.31\%) and integration (36.58\%). Test (25.96\%) and lint (27.84\%) also show notable prevalence. By contrast, analyze remains lower (15.87\%), and sync is rare (5.08\%). These phase differences indicate that caching is primarily deployed in phases where repeated dependency installation and artifact reuse is common (e.g., build and test), whereas jobs focused on synchronization or maintenance are less likely to benefit from caching.

\begin{table}[]
\centering
\caption{Prevalence of caching across different CI/CD job types.}
\label{tab:cicd_phase}
\begin{tabular}{lrrr}
\toprule
\multicolumn{1}{c}{\textbf{\shortstack{CI/CD job\\ types}}} & 
\multicolumn{1}{c}{\textbf{\shortstack{\# of jobs \\ with caching}}} & 
\multicolumn{1}{c}{\textbf{Total jobs}} & 
\multicolumn{1}{c}{\textbf{\shortstack{Caching\\ prevalence (\%)}}} \\ 
\midrule
build & 247 & 662 & 37.31\% \\ 
integration & 15 & 41 & 36.58\% \\ 
lint & 22 & 79 & 27.84\% \\ 
test & 141 & 543 & 25.96\% \\ 
release & 76 & 332 & 22.89\% \\ 
linux & 6 & 29 & 20.69\% \\ 
analyze & 20 & 126 & 15.87\% \\ 
sync & 3 & 59 & 5.08\% \\ 
other & 51 & 457 & 11.16\% \\ 
\midrule
\textbf{Total} & \textbf{581} & \textbf{2,328} & \textbf{24.95\%} \\ 
\bottomrule
\end{tabular}
\end{table}

\vspace{0.2cm}

\noindent \textbf{Observation 1.5: \textit{Repositories adopt caching mostly through three different caching strategies.}} We derived a taxonomy of cache types used in GHA workflows: 
\begin{enumerate}
    \item [-] \textbf{Explicit caching:} Caching is configured through an explicit cache step, using \texttt{actions/cache} where developers provide cache paths and keys. We define this  “explicit” because caching is enabled only when developers declare a cache step and specify its inputs, which offers flexibility to cache arbitrary directories (e.g., dependency folders, build outputs, tool caches), but requires the most manual configuration. 
    \item [-] \textbf{Package manager caching:} Caching is enabled through language/tool setup actions (e.g., \texttt{setup-go, setup-node, setup-python}) that either expose a caching option or provide caching behavior as part of the setup interface. We separate this from explicit cache actions because the developer’s configuration is mediated by the setup action, and the cached content is typically tied to the dependency management conventions of that ecosystem.
    \item [-] \textbf{Docker layer caching:} Caching is configured through Docker build tooling (e.g., docker/build-push-action) to reuse previously built layers. This is distinct from the two above mechanisms because the cached units are Docker image layers, and the configuration uses the Docker caching interface rather than file-path caching.
    \item [-] \textbf{Others:} Any caching configuration that does not fall into the above categories.
\end{enumerate}
\vspace{0.2cm}

\begin{table}[]
\centering
\caption{Distribution of caching strategies across CI/CD job types. Counts represent step-level caching configurations extracted from workflow files; a single job may contribute multiple times due to multiple caching steps.}
\label{tab:cache_type_phase}
\resizebox{\linewidth}{!}{%
\begin{tabular}{lrrrrr}
\toprule
\multicolumn{1}{c}{\textbf{\shortstack{CI/CD job\\ types}}} & 
\multicolumn{1}{c}{\textbf{\shortstack{Explicit\\ Caching}}} & 
\multicolumn{1}{c}{\textbf{\shortstack{Package Manager\\ Caching}}} & 
\multicolumn{1}{c}{\textbf{\shortstack{Docker Layer\\ Caching}}} & 
\multicolumn{1}{c}{\textbf{Others}} & 
\multicolumn{1}{c}{\textbf{Total}} \\ 
\midrule
build & 282 (73.1\%) & 71 (18.4\%) & 1 (0.3\%) & 32 (8.3\%) & 386 \\ 
test & 178 (75.4\%) & 53 (22.5\%) & 0 (0.0\%) & 5 (2.1\%) & 236 \\ 
release & 41 (41.0\%) & 59 (59.0\%) & 0 (0.0\%) & 0 (0.0\%) & 100 \\ 
other & 75 (80.6\%) & 17 (18.3\%) & 1 (1.1\%) & 0 (0.0\%) & 93 \\ 
analyze & 23 (74.2\%) & 6 (19.4\%) & 0 (0.0\%) & 2 (6.5\%) & 31 \\ 
lint & 19 (70.4\%) & 8 (29.6\%) & 0 (0.0\%) & 0 (0.0\%) & 27 \\ 
integration & 10 (66.7\%) & 5 (33.3\%) & 0 (0.0\%) & 0 (0.0\%) & 15 \\ 
linux & 7 (100.0\%) & 0 (0.0\%) & 0 (0.0\%) & 0 (0.0\%) & 7 \\ 
sync & 2 (66.7\%) & 1 (33.3\%) & 0 (0.0\%) & 0 (0.0\%) & 3 \\ 
\midrule
\textbf{Total} & \textbf{637 (70.9\%)} & \textbf{220 (24.5\%)} & \textbf{2 (0.2\%)} & \textbf{39 (4.3\%)} & \textbf{898} \\ 
\bottomrule
\end{tabular}%
}
\end{table}

\noindent \textbf{Observation 1.6: \textit{Explicit caching strategies dominate at the step level.}} Table~\ref{tab:cache_type_phase} shows that explicit caching via \texttt{actions/cache} accounts for the majority of cache-related steps (70.9\%). Package-manager caching via \texttt{setup} actions is the second most common strategy (24.5\%). 

A plausible explanation for the dominance of explicit caching is its flexibility and configurability. Developers can explicitly specify cache paths and keys, which enables caching beyond dependencies (e.g., build outputs and tool caches).\footnote{https://dev.to/github/caching-dependencies-to-speed-up-workflows-in-github-actions-3efl} Moreover, \texttt{actions/cache} is extensively documented and commonly referenced in tutorials, blog posts, and Q\&A forums such as Stack Overflow\footnote{https://stackoverflow.com/questions/74401969/what-is-the-logic-in-using-the-restore-keys-field-in-the-github-cache-action}, lowering the barrier to adoption through readily reusable examples \citep{baltes2019usage}. Its long-standing and stable interface may further make it an attractive choice from a maintainability and backward-compatibility perspective, whereas caching support in \texttt{setup-*} actions is mediated by ecosystem-specific conventions and may change over time.\footnote{https://github.com/actions/setup-node} Finally, explicit caching is well aligned with iterative CI/CD optimization, as its finer-grained and more flexible configuration allows developers to iteratively refine cache keys and paths in response to cache misses and performance bottlenecks.\footnote{https://docs.github.com/en/actions/reference/workflows-and-actions/dependency-caching}

\vspace{0.2cm}



\begin{tcolorbox}[enhanced,width=4.7in,size=fbox,drop shadow southwest,sharp corners]

\textbf{RQ1 Summary:} Repositories that adopt caching are more active than non-adopters. PHP and Java exhibit the highest adoption rates of CI/CD caching. Cache-adopting repositories maintain a moderate number of workflow files and jobs. At the job level, caching is most prevalent in build, integration, lint, and test jobs. At the step level, developers implement caching mainly through explicit caching strategies, followed by package manager caching and Docker layer caching.
\end{tcolorbox}

\section{RQ2: \rqtwo}\label{sec:rq2}

\subsection{Motivation}


While RQ1 represents the current, static characteristics of cache-adopting repositories, it does not explain how these configurations evolve over time. In practice, caching is rarely static or a one-time setup. As projects evolve and dependencies change, developers must update cache keys, paths, and strategies to maintain effectiveness and reduce CI/CD time. 

This research question examines these dynamics by analyzing the evolution of caching configurations across different job types. Understanding these evolution patterns is important because they reveal which cache maintenance activities are most common, how frequently configuration updates occur, and whether these dynamics differ across job types (e.g., build vs. test). These insights benefit practitioners by setting realistic expectations about the maintenance efforts required to keep caching effective and stable, and benefit tool and platform designers by highlighting where improved defaults, diagnostics, or guidance could reduce repeated trial-and-error in cache maintenance.

\subsection{Approach} Our evolution analysis is based on 266 cache-adopted repositories identified in Section \ref{sec:dataset}. Across these repositories, we identify 2,494 cache-related maintenance activities. For each job type, we reconstruct its chronological cache maintenance history from workflow histories, allowing us to trace how caching configurations evolve over time.We represent each maintenance activity as one of the seven cache-related maintenance activity categories defined in Section~\ref{sec:background}.

We model these histories using a first-order Markov model \citep{gagniuc2017markov}. Following the job taxonomy defined in in Section \ref{sec:dataset}, we group jobs by CI/CD job type and construct a separate transition model for each job type. In each model, nodes represent cache-related maintenance activity categories, and edges represent observed transitions between consecutive activities. For each transition, we compute its transition probability and corresponding time-to-transition statistics (in days), which are reported in the appendix, and visualize the resulting transition graphs (e.g., Figure~\ref{fig:build_tran}). To focus on recurring transitions, we prune edges with transition probabilities below 0.05, thereby emphasizing stable and representative transition patterns, following Huizi et al.~\citeyear{hao2024empirical}.

\subsection{Results}

\begin{figure}[h!]
    \centering
    \includegraphics[width=1.0\linewidth]{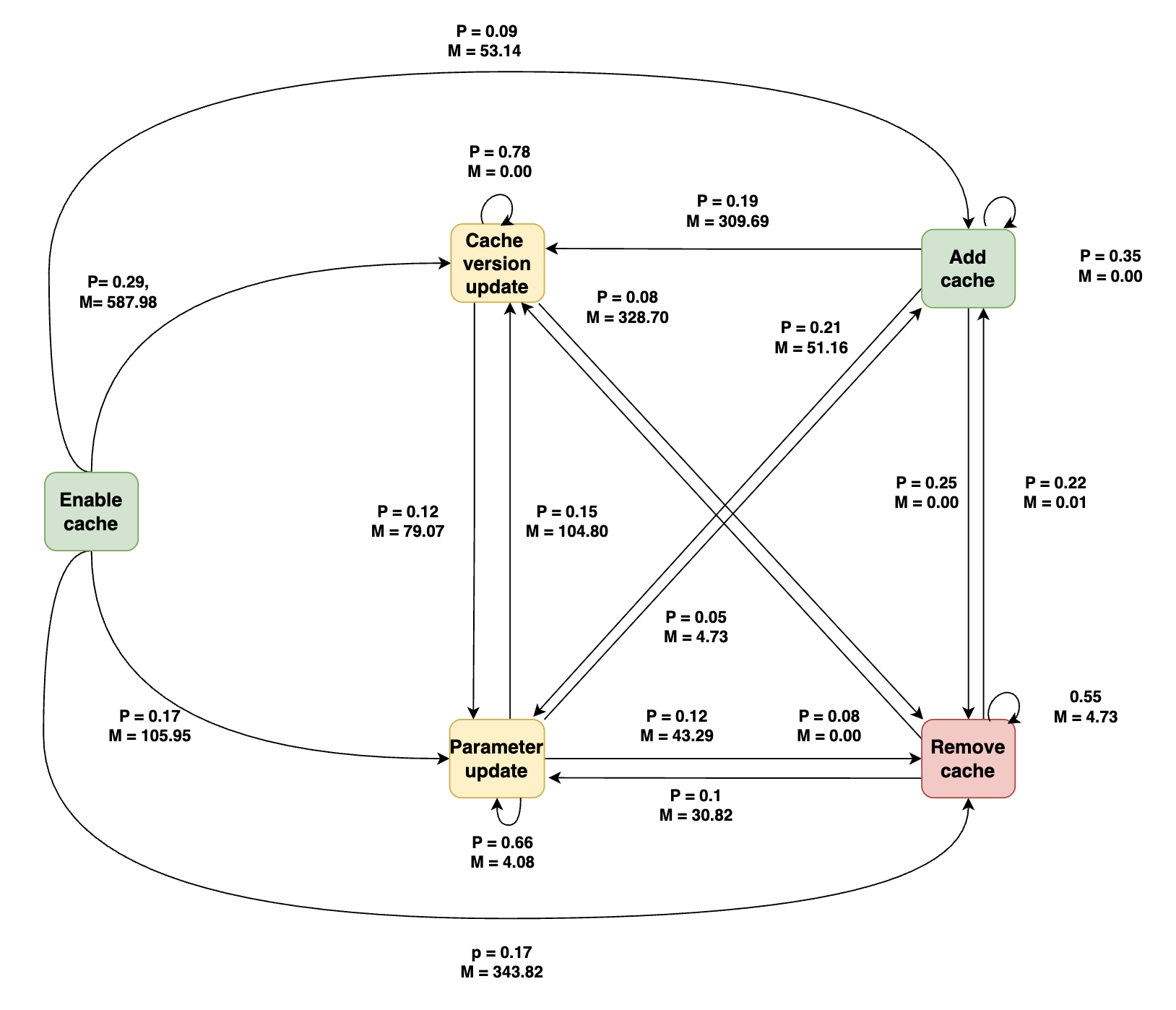}
    \caption{Cache-related maintenance activities in \textit{build jobs} after enabling caching. Edges report transition probability and time-to-transition in days.}
    \label{fig:build_tran}
\end{figure}

\begin{figure}[h!]
    \centering
    \includegraphics[width=1.0\linewidth]{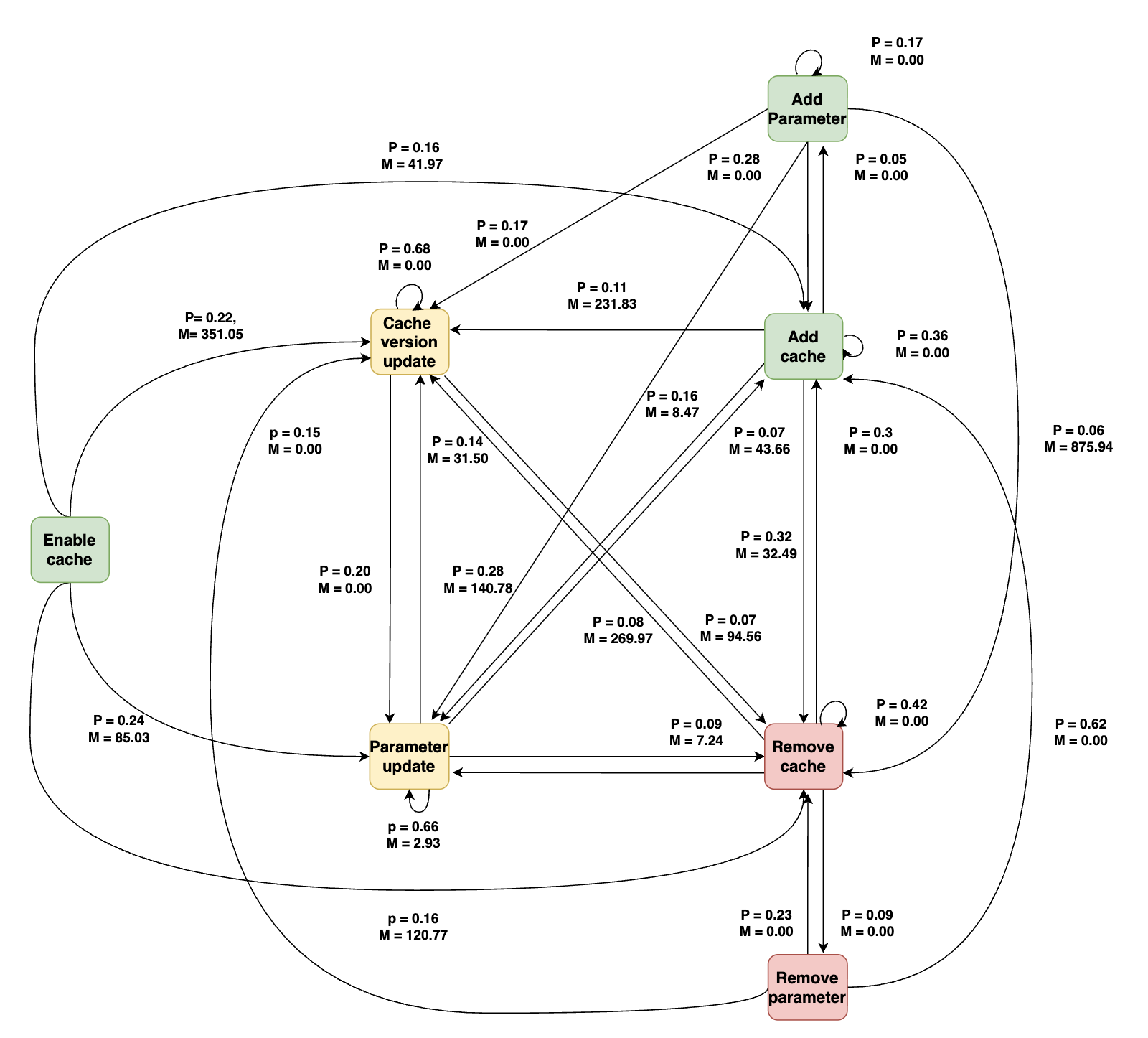}
    \caption{Cache-related maintenance activities in \textit{test jobs} after enabling caching. Edges report transition probability and time-to-transition in days.}
    \label{fig:test_tran}
\end{figure}

\begin{figure}[h!]
    \centering
    \includegraphics[width=1.0\linewidth]{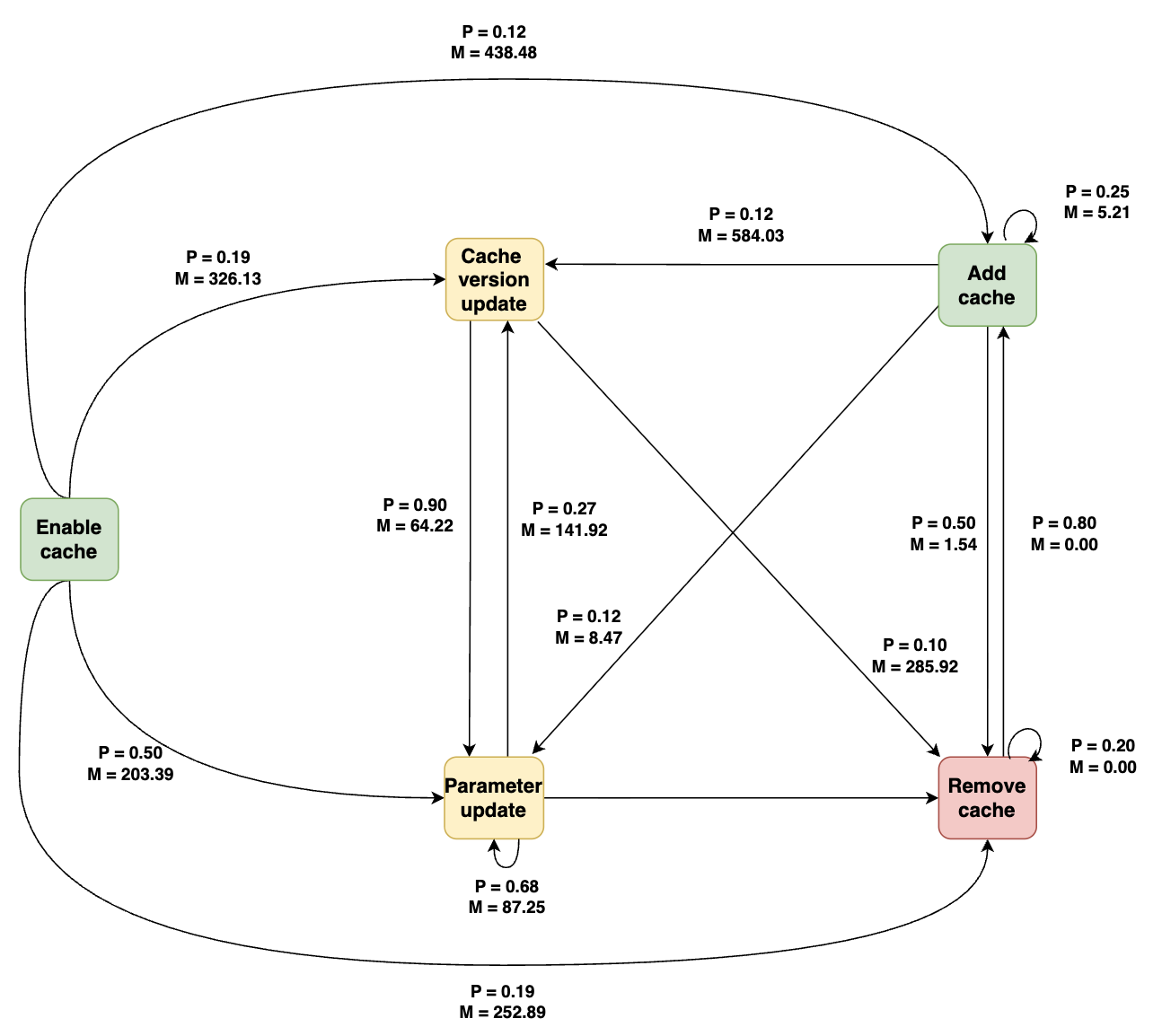}
    \caption{Cache-related maintenance activities in \textit{integration jobs} after enabling caching. Edges report transition probability and time-to-transition in days.}
    \label{fig:inte_trans}
\end{figure}

\begin{figure}[h!]
    \centering
    \includegraphics[width=1.0\linewidth]{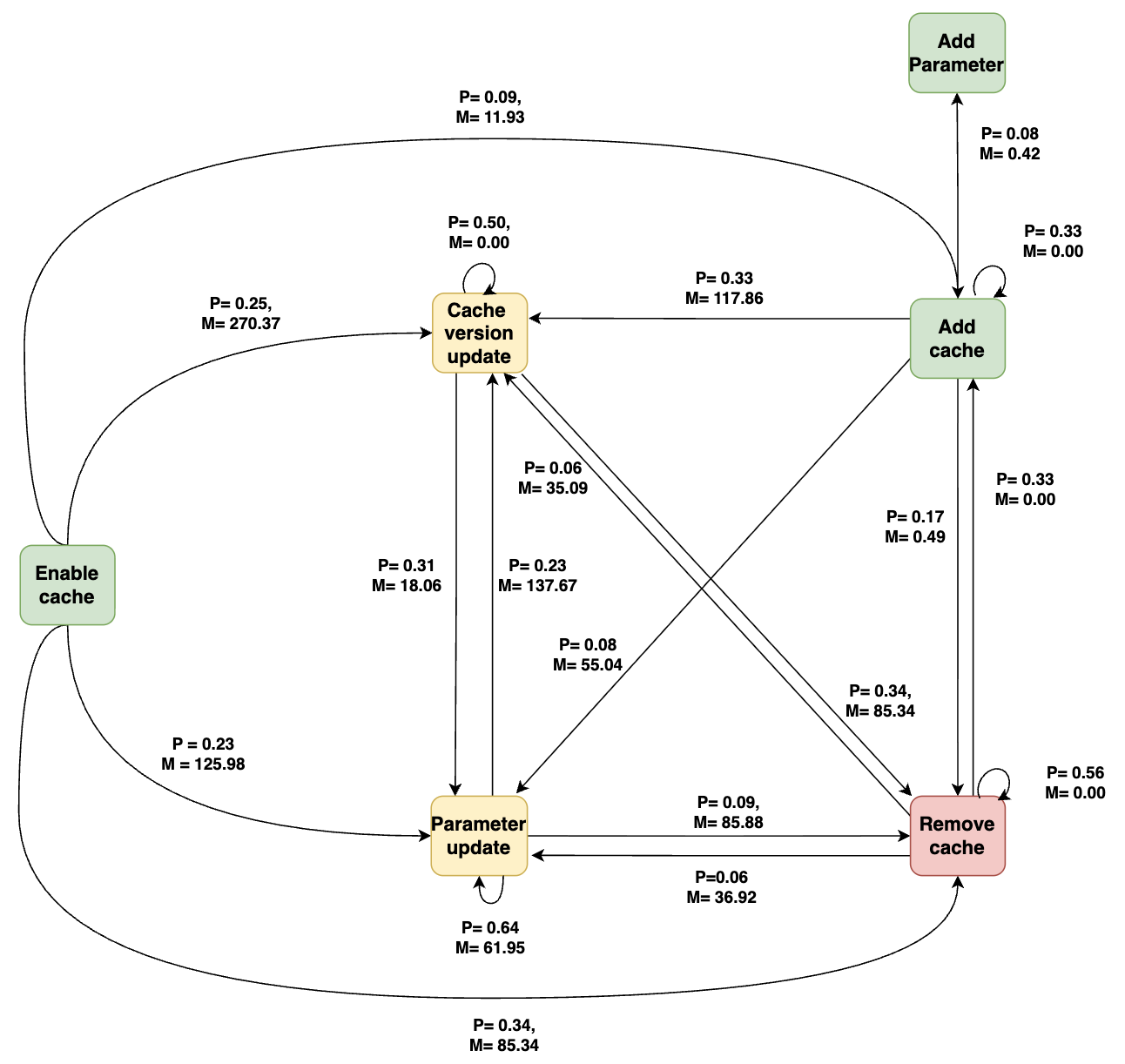}
    \caption{Cache-related maintenance activities in \textit{release jobs} after enabling caching. Edges report transition probability and time-to-transition in days.}
    \label{fig:inte_release}
\end{figure}

\begin{figure}[h!]
    \centering
    \includegraphics[width=1.0\linewidth]{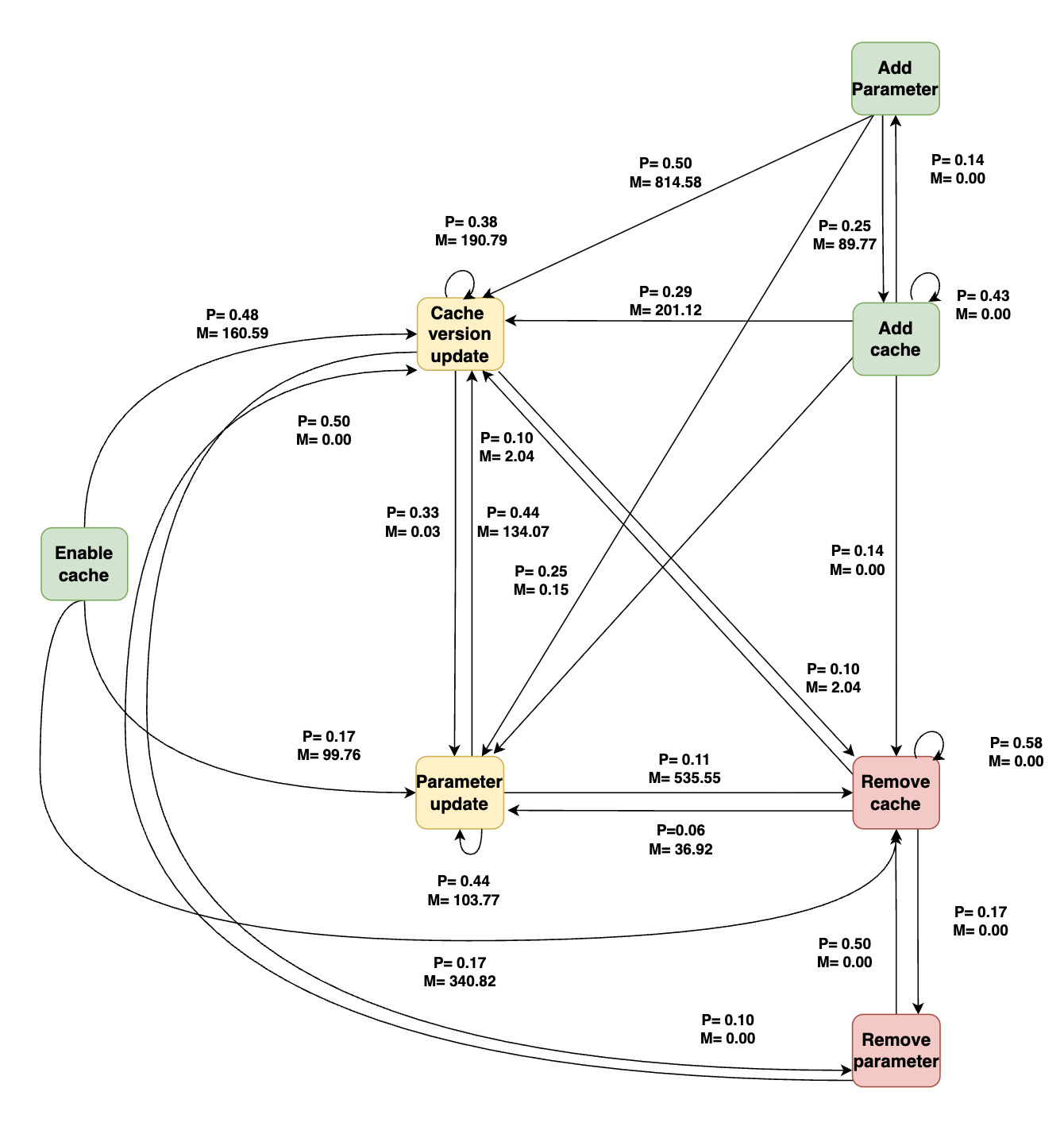}
    \caption{Cache-related maintenance activities in \textit{lint jobs} after enabling caching. Edges report transition probability and time-to-transition in days.}
    \label{fig:lint_trans}
\end{figure}

\begin{figure}[h!]
    \centering
    \includegraphics[width=1.0\linewidth]{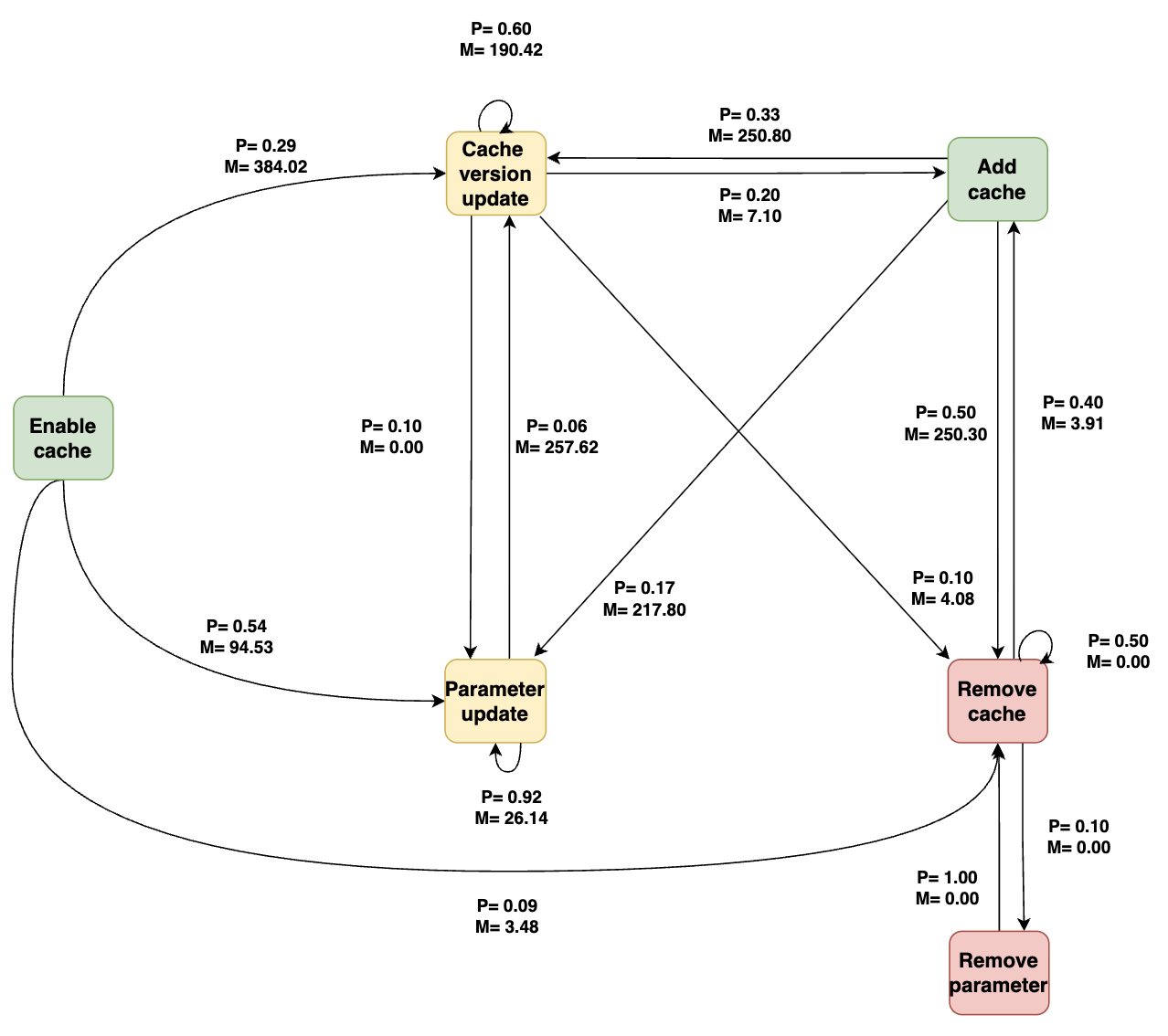}
    \caption{Cache-related maintenance activities in \textit{analyze jobs} after enabling caching. Edges report transition probability and time-to-transition in days.}
    \label{fig:tran_analyze}
\end{figure}

\begin{figure}[h!]
    \centering
    \includegraphics[width=1.0\linewidth]{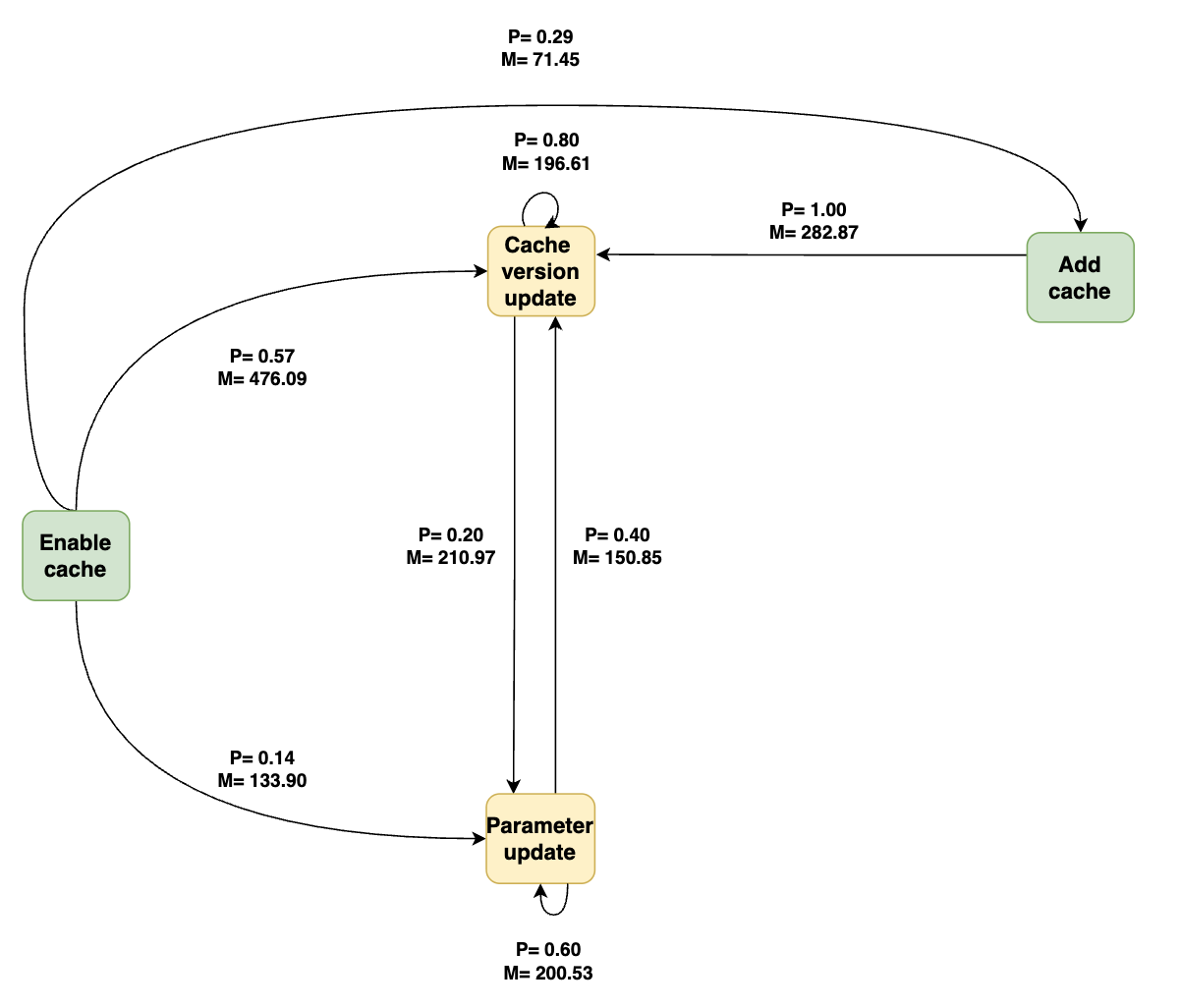}
    \caption{Cache-related maintenance activities in \textit{linux jobs} after enabling caching. Edges report transition probability and time-to-transition in days.}
    \label{fig:linux_trans}
\end{figure}


\begin{figure}[h!]
    \centering
    \includegraphics[width=1.0\linewidth]{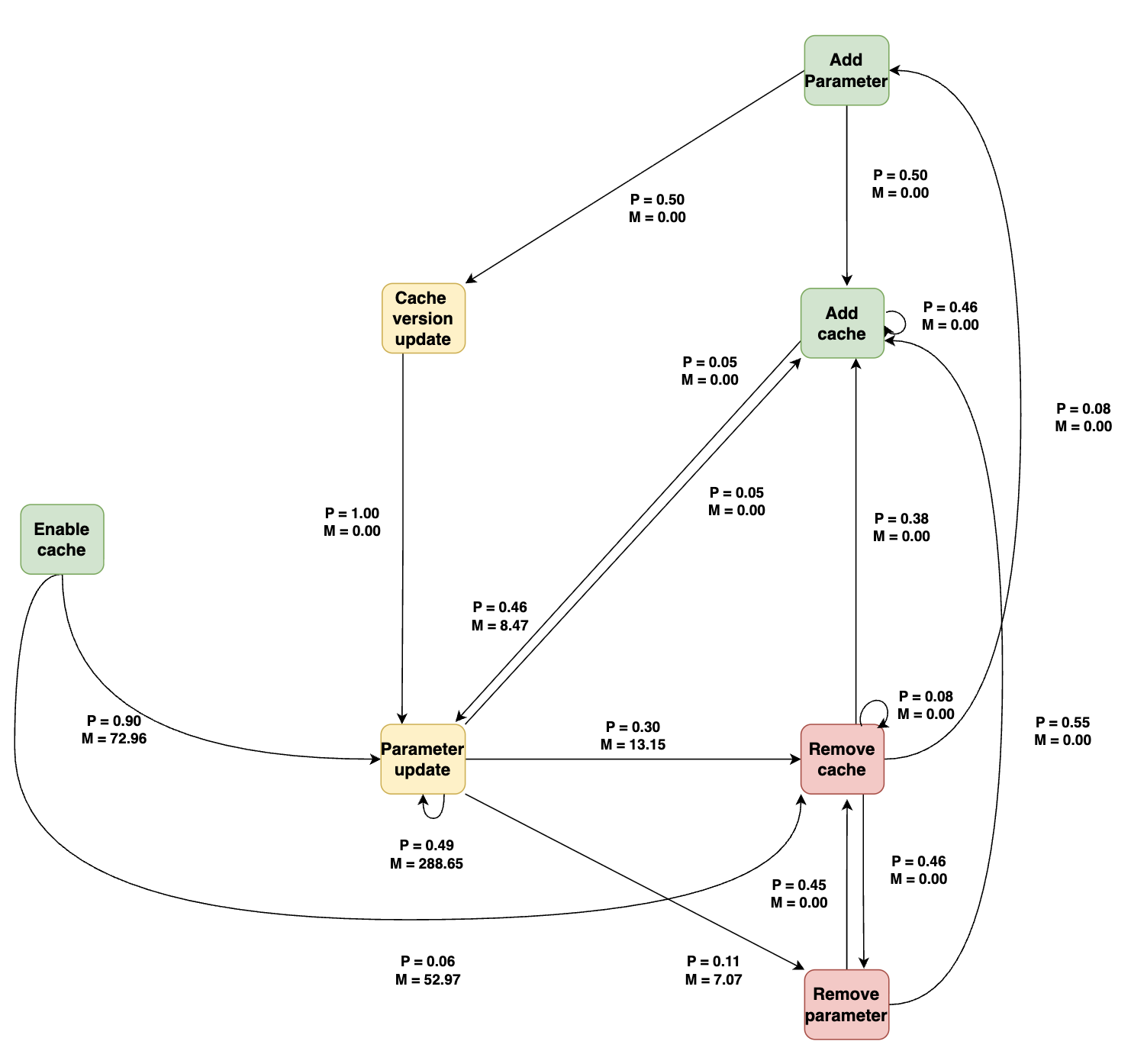} 
    \caption{Cache-related maintenance activities in \textit{sync jobs} after enabling caching. Edges report transition probability and time-to-transition in days.}
    \label{fig:trans_sync}
\end{figure}

\vspace{0.2cm}

In the following paragraphs, we describe how caching configurations evolve over time through cache maintenance activities across CI/CD job types. We begin by describing the overall cache evolution patterns visible in the transition plots, and then report the dominant transition behaviors and timing differences across job types. In Figure~\ref{fig:build_tran} to Figure~\ref{fig:trans_sync}, node colors distinguish the functional role of each cache-related maintenance activity. \textbf{Green} denotes cache introduction and extension (\textit{Enable cache}, \textit{Add cache}, \textit{Add parameter}), \textbf{yellow} denotes maintenance (\textit{Cache version update}, \textit{Parameter update}), and \textbf{red} denotes removal (\textit{Remove cache}, \textit{Remove parameter}). Throughout, we refer to the transition probabilities and the median time between consecutive activities. The complete evolution tables for all job types are provided in Appendix~\ref{sec:appen}.









\vspace{0.2cm}

\noindent \textbf{Observation 2.1:  \textit{Cache evolution across CI/CD job types follows an iterative pattern of enabling, updating, adding, and occasionally removing cache configurations.}} Across all job-type-specific transition graphs (Figure \ref{fig:build_tran}- \ref{fig:trans_sync}), cache evolution consistently begins with \textit{Enable cache}, which serves as the entry point of cache adoption. After this initial step, the evolution is dominated by maintenance-related states, particularly \textit{Cache version update} and \textit{Parameter update}, which appear across all job types. In addition to these maintenance activities, cache evolution also includes expansion and adjustment operations. All job types involve \textit{Add cache}, indicating that caching configurations are incrementally extended after initial adoption. Furthermore, several job types (build, test, integration, release etc.) include removal-related states (e.g., removing caches or parameters), suggesting that developers may roll back previous caching decisions. This suggests that cache configuration is an iterative process involving repeated updates and adjustments, rather than a single-pass setup.

\vspace{0.2cm}

\noindent \textbf{Observation 2.2: \textit{The structural complexity of cache evolution graphs differs across CI/CD job types.}} To characterize the complexity of cache evolution, we examine the structure of the transition graphs for each job type, including both the number of distinct cache-related maintenance activity states (nodes) and the transitions between them (edges). The transition graphs show clear variation across job types. Among all job types, \textit{Test} and \textit{Lint} jobs involve the largest number of distinct nodes after enabling cache. In contrast, \textit{Linux} jobs show the smallest evolution space with only four nodes (\textit{Enable cache}, \textit{Add cache}, \textit{Cache version update}, \textit{Parameter update}). \textit{Build} and \textit{Integration} jobs each include five different nodes, indicating a broader range of cache-related activities. This indicates that cache evolution differs across job types, with some job types involving a broader set of cache maintenance activities than others.




\vspace{0.2cm}


\noindent \textbf{Observation 2.3: \textit{Cache maintenance in several CI/CD job types is dominated by repeated self-loop transitions.}} While Observation 2.1 shows that cache evolution generally involves enabling, updating, adding, and sometimes removing cache configurations, the transition tables further show that the most prevalent maintenance pattern in several job types is repeated self-loop behavior. In particular, across \textit{build}, \textit{test}, \textit{integration}, and \textit{release} jobs, the dominant transitions are often $P\_up \rightarrow P\_up$ and $C\_up \rightarrow C\_up$, indicating repeated parameter tuning and repeated cache-version maintenance within the same state. For example, in \textit{build} jobs, $C\_up \rightarrow C\_up$ accounts for 77.63\% of outgoing transitions from \textit{C\_up}, with a median time of 0 days (Table~\ref{tab:build_trans_short}, Figure~\ref{fig:build_tran}). This indicates that cache maintenance is rarely a static implementation. Once a developer begins tuning parameters and updating cache versions, they tend to continue doing so in a sequence of related commits. This suggests that maintaining a cache is an iterative process requiring sustained attention rather than a single action.



\vspace{0.2cm}

\vspace{0.2cm}


\noindent \textbf{Observation 2.4: \textit{Parameter updates occur earlier than cache version updates, but the delay differs across job types.}} Shortly after enabling cache ($EC$), developers engage in parameter tuning ($EC \rightarrow P\_up$) relatively quickly, with a median of 105.93 days for Build jobs (Table \ref{tab:build_trans_short}, Fig. \ref{fig:build_tran}). However, the maintenance activity of updating the cache version ($EC \rightarrow C\_up$) occurs after a significantly longer period. It has a median of 270.37 days in Release jobs (Table \ref{tab:release_trans_short}, Fig. \ref{fig:inte_release}) and 384.02 days in Analyze jobs (Table \ref{tab:analyze_trans_short}, Fig \ref{fig:tran_analyze}). This indicates that once a stable cache version is adopted, developers are reluctant to change it.  While they may tweak how the cache is saved through parameters, they often let the core cache version remain unchanged for a much longer period.

\vspace{0.2cm}

\noindent \textbf{Observation 2.5: \textit{Cache removal is often part of an immediate replacement.}} In build jobs, the transition from removing a cache to immediately adding a new one ($C\_rm \rightarrow C\_add$) occurs with 21.84\% probability and a median of 0.01 days (Table \ref{tab:build_trans_short}, Fig. \ref{fig:build_tran}). Similarly, in Test jobs, this occurs with 30.19\% probability and a median of 0.00 days (Table \ref{tab:test_trans_short}, Fig. \ref{fig:test_tran}).  The near-zero median time indicates that developers are not permanently removing caches; they are replacing an old configuration with a new one. This suggests that the initial cache setup was insufficient, prompting a switch to a different strategy rather than abandoning caching altogether.

\vspace{0.2cm}

\noindent \textbf{Observation 2.6: \textit{In build and test jobs, cache evolution is characterized by frequent and rapid repeated parameter tuning followed by later expansion through additional caching.}} The $P\_up \rightarrow P\_up$ activity is dominant in both job types. In \textit{build} jobs, it occurs with 65.53\% probability and a median of 4.08 days (Table~\ref{tab:build_trans_short}, Figure~\ref{fig:build_tran}). In \textit{test} jobs, it occurs with 66.05\% probability and a median of 2.93 days (Table~\ref{tab:test_trans_short}, Figure~\ref{fig:test_tran}). In the same job types, cache expansion through $EC \rightarrow C\_add$ occurs later, with median times of 53.14 days in \textit{build} jobs and 41.97 days in \textit{test} jobs. This suggests a two-stage evolution pattern in build and test jobs. Developers first repeatedly refine cache-related parameters soon after adoption, likely to stabilize the initial setup, and only later extend the caching strategy by adding new caches.

\vspace{0.2cm}

\noindent \textbf{Observation 2.7: \textit{Integration and release jobs exhibit long intervals between cache maintenance activities.}} In contrast to build and test jobs, the maintenance activity of updating parameters ($P\_up \rightarrow P\_up$) happens much less often. The median time between these updates is 87.25 days for Integration jobs (Table \ref{tab:integration_trans_short}, Fig. \ref{fig:inte_trans}) and 61.95 days for Release jobs (Table \ref{tab:release_trans_short}, Fig, \ref{fig:inte_release}). Furthermore, the time from enabling cache to the first parameter update ($EC \rightarrow P\_up$) in Integration jobs has a median of 203.39 days. This indicates that cache configurations in these job types are occasionally modified. Unlike Build or Test jobs where changes happen frequently, developers do not often update the cache settings for Integration and Release jobs once they are established.


\begin{tcolorbox}[enhanced,width=4.7in,size=fbox,drop shadow southwest,sharp corners]
\textbf{RQ2 Summary: Cache evolution is an iterative and structurally complex process. Across all job types, cache evolution consistently involves maintaining existing configurations through parameter updates and cache version updates, often accompanied by incremental additions and, in some cases, removal of cache-related configurations. Cache configurations are repeatedly refined rather than modified in a single pass. The structural complexity of this evolution varies across job types: Test and Build jobs involve richer transition structures and more frequent maintenance, whereas Integration and Release jobs exhibit simpler structures and substantially longer intervals between maintenance activities.}
\end{tcolorbox}


\section{RQ3: \rqthree}\label{sec:rq3}
\subsection{Motivation}
RQ2 examines how caching configurations evolve over time across different CI/CD job types by modeling sequences of cache-related maintenance activities as transition graphs. We observed rapid, repeated parameter updates in build and test jobs, whereas maintenance activities in other job types (e.g., release) occur less frequently and over longer intervals. However, these temporal patterns alone do not explain why developers make these modifications. For example, a rapid sequence of parameter updates can reflect deliberate performance refinement, but it can also indicate repeated attempts to resolve cache-related issues. Similarly, the long delay before a cache version update can reflect continued reliance on an existing configuration, or it can indicate that dependency maintenance is postponed until a warning or automated alert triggers action. To move from describing \emph{what} changes and \emph{when} they occur to explaining \emph{why} they occur, RQ3 investigates the underlying drivers of cache-related modifications in GHA workflows, using commit/PR context and PR author information (human vs.\ bot) to interpret the intent behind the patterns observed in RQ2.

\subsection{Approach}
We answer RQ3 by linking the quantitative cache-related maintenance transition patterns observed in RQ2 to qualitative explanations derived from the developer context. We proceed in two steps. First, we summarize transitions by their target state to identify which types of cache-related modifications are dominant and therefore require explanation (Table \ref{tab:rq3}). This table provides a quantitative basis for RQ3 and is discussed in the Section \ref{sec:rq3_result} to justify our qualitative focus on particular targets. Second, we perform a qualitative analysis on a statistically significant random sample of transitions for each target state, using a 90\% confidence level and a 10\% margin of error \citep{almulla2025understanding}. Specifically, we manually analyzed 86 (\textit{cache version update} (C\_up)), 87 (\textit{parameter update} (P\_up)), 77 (\textit{adding cache} (C\_add)), 28 (\textit{adding parameter} (P\_add)), 79 (\textit{removing cache} (C\_rm)), and 32 (\textit{removing parameter} (P\_rm)) sampled transitions.

For each sampled transition, we examined the workflow file change that introduced the cache-related modification, including the workflow YAML before and after the edit. To infer the rationale behind the change, we examined the associated commit message and, when available, the linked pull request title and description. We identified the initiator of each change from the commit or pull request author and classified it as human or bot using author identity and automation indicators such as Dependabot\footnote{https://github.com/dependabot} and Renovate\footnote{https://github.com/renovatebot/renovate}. Finally, we summarize the driver categories and their frequencies.


\subsection{Results}

\label{sec:rq3_result}

\begin{table}[]
\centering
\caption{Statistics of cache-related maintenance transitions, aggregated by target state. For each target cache maintenance activity, we report transition counts from different preceding activities together with the corresponding transition time statistics (in days).}
\label{tab:rq3}
\resizebox{\linewidth}{!}{%
\begin{tabular}{llrrrrrrrr}
\toprule
\textbf{From} & \textbf{To} & \textbf{Count} & \textbf{Total} & \textbf{Avg} & \textbf{Med} & \textbf{Std} & \textbf{Min} & \textbf{Max} & \textbf{\%} \\ 
\midrule
Cache version update & \multirow{7}{*}{\textbf{Cache version update}} & 378 & \multirow{7}{*}{745} & 70.60 & 0.00 & 149.79 & 0.00 & 713.59 & 51.85 \\ 
Enable cache &  & 168 &  & 543.33 & 384.13 & 447.09 & 0.01 & 1615.93 & 23.05 \\ 
Parameter update &  & 104 &  & 172.52 & 104.80 & 241.06 & 0.00 & 1366.00 & 14.27 \\ 
Adding cache &  & 44 &  & 298.89 & 231.83 & 310.83 & 0.00 & 1294.85 & 6.04 \\ 
Removing cache &  & 18 &  & 210.66 & 95.11 & 334.78 & 0.00 & 1374.41 & 2.47 \\ 
Adding parameter &  & 12 &  & 179.04 & 0.00 & 392.71 & 0.00 & 1366.00 & 1.65 \\ 
Removing parameter &  & 5 &  & 0.00 & 0.00 & 0.00 & 0.00 & 0.00 & 0.69 \\ 
\midrule
Parameter update & \multirow{7}{*}{\textbf{Parameter update}} & 497 & \multirow{7}{*}{854} & 75.36 & 4.29 & 148.32 & 0.00 & 986.95 & 58.20 \\ 
Enable cache &  & 166 &  & 207.31 & 115.90 & 260.34 & 0.00 & 981.44 & 19.44 \\ 
Cache version update &  & 92 &  & 95.24 & 0.23 & 136.98 & 0.00 & 592.69 & 10.77 \\ 
Adding cache &  & 66 &  & 106.08 & 18.14 & 176.34 & 0.00 & 973.79 & 7.73 \\ 
Removing cache &  & 23 &  & 59.31 & 14.83 & 132.34 & 0.00 & 584.88 & 2.69 \\ 
Adding parameter &  & 8 &  & 100.00 & 3.23 & 220.22 & 0.00 & 628.32 & 0.94 \\ 
Removing parameter &  & 2 &  & 117.53 & 117.53 & 166.21 & 0.00 & 235.06 & 0.23 \\ 
\midrule
Adding cache & \multirow{7}{*}{\textbf{Adding cache}} & 129 & \multirow{7}{*}{362} & 19.54 & 0.00 & 76.00 & 0.00 & 463.22 & 35.64 \\ 
Removing cache &  & 80 &  & 49.43 & 0.00 & 121.76 & 0.00 & 538.67 & 22.10 \\ 
Enable cache &  & 72 &  & 158.53 & 46.39 & 212.73 & 0.00 & 962.46 & 19.89 \\ 
Parameter update &  & 39 &  & 70.13 & 10.86 & 153.92 & 0.00 & 903.97 & 10.77 \\ 
Cache version update &  & 18 &  & 54.97 & 16.83 & 96.33 & 0.00 & 380.01 & 4.97 \\ 
Removing parameter &  & 14 &  & 0.00 & 0.00 & 0.00 & 0.00 & 0.00 & 3.87 \\ 
Adding parameter &  & 10 &  & 0.00 & 0.00 & 0.00 & 0.00 & 0.00 & 2.76 \\ 
\midrule
Adding cache & \multirow{7}{*}{\textbf{Adding parameter}} & 11 & \multirow{7}{*}{47} & 0.04 & 0.00 & 0.13 & 0.00 & 0.42 & 28.21 \\ 
Parameter update &  & 9 &  & 96.68 & 4.92 & 253.28 & 0.00 & 769.43 & 23.08 \\ 
Cache version update &  & 6 &  & 2.20 & 0.00 & 5.38 & 0.00 & 13.18 & 15.38 \\ 
Enable cache &  & 5 &  & 1.01 & 0.00 & 2.27 & 0.00 & 5.07 & 12.82 \\ 
Adding parameter &  & 4 &  & 0.00 & 0.00 & 0.00 & 0.00 & 0.00 & 10.26 \\ 
Removing cache &  & 3 &  & 57.74 & 0.00 & 100.02 & 0.00 & 173.23 & 7.69 \\ 
Removing parameter &  & 1 &  & 94.21 & 94.21 & 0.00 & 94.21 & 94.21 & 2.56 \\ 
\midrule
Removing cache & \multirow{7}{*}{\textbf{Removing cache}} & 117 & \multirow{7}{*}{439} & 21.59 & 0.00 & 79.25 & 0.00 & 448.01 & 26.65 \\ 
Enable cache &  & 111 &  & 293.62 & 129.68 & 362.73 & 0.00 & 1514.81 & 25.28 \\ 
Adding cache &  & 76 &  & 110.33 & 4.25 & 167.29 & 0.00 & 750.21 & 17.31 \\ 
Parameter update &  & 69 &  & 168.77 & 43.29 & 270.95 & 0.00 & 1154.81 & 15.72 \\ 
Cache version update &  & 45 &  & 201.85 & 125.65 & 205.24 & 0.00 & 707.67 & 10.25 \\ 
Removing parameter &  & 20 &  & 0.00 & 0.00 & 0.00 & 0.00 & 0.00 & 4.56 \\ 
Adding parameter &  & 1 &  & 875.94 & 875.94 & 0.00 & 875.94 & 875.94 & 0.23 \\ 
\midrule
Removing cache & \multirow{7}{*}{\textbf{Removing parameter}} & 26 & \multirow{7}{*}{47} & 7.36 & 0.00 & 37.52 & 0.00 & 191.32 & 54.17 \\ 
Parameter update &  & 10 &  & 16.26 & 7.07 & 24.39 & 0.00 & 70.85 & 20.83 \\ 
Adding cache &  & 3 &  & 1.09 & 0.00 & 1.89 & 0.00 & 3.27 & 6.25 \\ 
Cache version update &  & 3 &  & 57.31 & 0.00 & 99.26 & 0.00 & 171.93 & 6.25 \\ 
Adding parameter &  & 2 &  & 43.47 & 43.47 & 61.48 & 0.00 & 86.94 & 4.17 \\ 
Enable cache &  & 2 &  & 432.13 & 432.13 & 433.11 & 125.88 & 738.39 & 4.17 \\ 
Removing parameter &  & 2 &  & 0.00 & 0.00 & 0.00 & 0.00 & 0.00 & 4.17 \\ 
\bottomrule
\end{tabular}%
}
\end{table}

\begin{table}[h]
\centering
\caption{Taxonomy of drivers for cache-related modifications.}
\label{tab:caching_taxonomy}
\small
\begin{tabularx}{\linewidth}{@{}llXl@{}}
\toprule
\textbf{Driver} & \textbf{Source} & \textbf{Definition (D) \& Example (E)} & \textbf{Freq.} \\ 
\midrule
Update dependencies & Human & \textbf{D:} Developers update, upgrade, or downgrade dependencies of caching configurations. \newline \textit{\textbf{E:} 'Update Go dependencies'} & 91 \\ \midrule
Improve cache & Human & \textbf{D:} Changes to improve caching performance. \newline \textit{\textbf{E:} 'improved github workflows'} & 47 \\ \midrule
Bot alert & Bots & \textbf{D:} Dependency updates triggered by bot alerts. \newline \textit{\textbf{E:} 'Bump actions/cache from 4.0.0 to 4.0.2'} & 40 \\ \midrule
Extend cache & Human & \textbf{D:} Adding additional caching configuration to the yaml file (e.g., expand support). \newline \textit{\textbf{E:} 'use multiple jobs'} & 28 \\ \midrule
Fix caching issues & Human & \textbf{D:} Developers address/fix issues related to caching. \newline \textit{\textbf{E:} 'Fix cmake.yml cache issues'} & 23 \\ \midrule
Migration & Human & \textbf{D:} Migrate to a different workflow (e.g., pnpm). \newline \textit{\textbf{E:} 'migrate to pnpm'} & 14 \\ \midrule
Clean up & Human & \textbf{D:} Cleanup, reconfigure, or refactor yaml files. \newline \textit{\textbf{E:} 'Simplify pipeline'} & 14 \\ \midrule
Remove tests & Human & \textbf{D:} Removal of specific jobs/tests after a period of time. \newline \textit{\textbf{E:} 'Remove legacy tests and style guide'} & 13 \\ \midrule
Fix security issues & Human/Bot & \textbf{D:} Address security concerns by updating configurations. \newline \textit{\textbf{E:} 'ci: Harden GitHub Actions'} & 6 \\ \midrule
Disable cache & Human & \textbf{D:} Developers explicitly disable caching. \newline \textit{\textbf{E:} 'Removed caching of compiled contracts'} & 4 \\ \midrule
Fix deprecated ver. & Human & \textbf{D:} Fixing nearly deprecated version or addressing warnings. \newline \textit{\textbf{E:} 'Change GitHub Action version to address deprecation'} & 2 \\ \midrule
Other & - & Reason not stated or unclear in commit/PR. & 107 \\ 
\bottomrule
\end{tabularx}
\end{table}

\begin{table}[!htbp]
\centering
\caption{Summary of caching change targets, extended actions, and underlying reasons.}
\label{tab:rq3_table_1}
\scriptsize
\renewcommand{\arraystretch}{1.2}
\begin{tabularx}{\linewidth}{@{}llXl@{}}
\toprule
\textbf{Target} & \textbf{Action} & \textbf{Drivers (Count)} & \textbf{Total} \\ 
\midrule

\multirow{3}{*}{\textbf{Cache version update}} 
 & Update & Fix security issues (1) & 1 \\
\cmidrule(lr){2-4}
 & Upgrade & Update dependencies (41) \newline Bot alert (40) \newline Fix deprecated versions (2) \newline Other (1) & 84 \\
\cmidrule(lr){2-4}
 & Downgrade & Update dependencies (1) & 1 \\
\midrule

\multirow{3}{*}{\textbf{Parameter update}} 
 & Update & Fix caching issues (9) \newline Update dependencies (7) \newline Improve cache (1) \newline Fix security issues (1) \newline Migration (1) \newline Other (22) & 41 \\
\cmidrule(lr){2-4}
 & Upgrade & Update dependencies (35) \newline Fix caching issues (5) \newline Fix security issues (1) \newline Clean up (1) \newline Other (1) & 43 \\
\cmidrule(lr){2-4}
 & Downgrade & Fix deprecated versions (2) \newline Update dependencies (1) & 3 \\
\midrule

\textbf{Adding cache} & Add & Extend cache (28) \newline Improve cache (17) \newline Fix caching issues (4) \newline Fix security issues (3) \newline Migration (2) \newline Other (23) & 77 \\
\midrule

\textbf{Removing cache} & Remove & Improve cache (14) \newline Remove tests (13) \newline Clean up (10) \newline Migration (4) \newline Disable cache (4) \newline Fix caching issues (1) \newline Other (33) & 79 \\
\midrule

\textbf{Adding parameter} & Add & Improve cache (12) \newline Update dependencies (4) \newline Fix caching issues (2) \newline Migration (2) \newline Clean up (1) \newline Other (7) & 28 \\
\midrule

\textbf{Removing parameter} & Remove & Migration (5) \newline Update dependencies (2) \newline Clean up (2) \newline Improve cache (1) \newline Other (22) & 32 \\

\bottomrule
\end{tabularx}
\end{table}

\begin{table}[h]
\centering
\caption{Frequency distribution of drivers across different cache modification targets.}
\label{tab:my-table}
\resizebox{\linewidth}{!}{%
\begin{tabular}{lrrrrrrr}
\toprule
\textbf{Categories} & 
\multicolumn{1}{c}{\rotatebox{45}{\textbf{Adding Cache}}} & 
\multicolumn{1}{c}{\rotatebox{45}{\textbf{Adding Param}}} & 
\multicolumn{1}{c}{\rotatebox{45}{\textbf{Cache Ver. Upd}}} & 
\multicolumn{1}{c}{\rotatebox{45}{\textbf{Removing Cache}}} & 
\multicolumn{1}{c}{\rotatebox{45}{\textbf{Removing Param}}} & 
\multicolumn{1}{c}{\rotatebox{45}{\textbf{Param Update}}} & 
\textbf{Total} \\ 
\midrule
Update dependencies & 0 (0.0\%) & 4 (4.4\%) & 42 (46.2\%) & 0 (0.0\%) & 2 (2.2\%) & 43 (47.3\%) & 91 \\ 
Improve cache & 17 (36.2\%) & 12 (25.5\%) & 0 (0.0\%) & 14 (29.8\%) & 1 (2.1\%) & 3 (6.4\%) & 47 \\ 
Bot alert & 0 (0.0\%) & 0 (0.0\%) & 40 (100.0\%) & 0 (0.0\%) & 0 (0.0\%) & 0 (0.0\%) & 40 \\ 
Extend cache & 28 (100.0\%) & 0 (0.0\%) & 0 (0.0\%) & 0 (0.0\%) & 0 (0.0\%) & 0 (0.0\%) & 28 \\ 
Fix caching issues & 4 (17.4\%) & 2 (8.7\%) & 0 (0.0\%) & 1 (4.3\%) & 0 (0.0\%) & 16 (69.6\%) & 23 \\ 
Clean up & 0 (0.0\%) & 1 (7.1\%) & 0 (0.0\%) & 10 (71.4\%) & 2 (14.3\%) & 1 (7.1\%) & 14 \\ 
Migration & 2 (14.3\%) & 2 (14.3\%) & 0 (0.0\%) & 4 (28.6\%) & 5 (35.7\%) & 1 (7.1\%) & 14 \\ 
Remove tests & 0 (0.0\%) & 0 (0.0\%) & 0 (0.0\%) & 13 (100.0\%) & 0 (0.0\%) & 0 (0.0\%) & 13 \\ 
Fix security issues & 3 (50.0\%) & 0 (0.0\%) & 1 (16.7\%) & 0 (0.0\%) & 0 (0.0\%) & 2 (33.3\%) & 6 \\ 
Disable cache & 0 (0.0\%) & 0 (0.0\%) & 0 (0.0\%) & 4 (100.0\%) & 0 (0.0\%) & 0 (0.0\%) & 4 \\ 
Fix deprecated ver. & 0 (0.0\%) & 0 (0.0\%) & 2 (100.0\%) & 0 (0.0\%) & 0 (0.0\%) & 0 (0.0\%) & 2 \\ 
Other & 23 (21.5\%) & 7 (6.5\%) & 1 (0.9\%) & 33 (30.8\%) & 22 (20.6\%) & 21 (19.6\%) & 107 \\ 
\bottomrule
\end{tabular}%
}
\end{table}

We present the findings of RQ3 in a structured manner. We first use Table~\ref{tab:rq3} to identify the most frequently observed target states in the cache-related transition that require explanation. We then interpret these target states using the qualitative driver taxonomy and initiator information in Tables~\ref{tab:caching_taxonomy}, \ref{tab:rq3_table_1}, and \ref{tab:my-table}. Table~\ref{tab:rq3_table_1} links each target state to its specific modification action and associated drivers, while Table~\ref{tab:my-table} shows how driver categories are distributed across different cache modification targets. The \textit{Other} category (N=107) in Table \ref{tab:caching_taxonomy} is excluded from the discussion because it concerns cases in which the commit/PR text does not provide a clear rationale for cache-related modifications.









\vspace{0.2cm}

\noindent \textbf{Observation 3.1: \textit{Overall, most cache-related modifications are driven by dependency updates, improving cache, and bot alerts.}} Table \ref{tab:caching_taxonomy} shows that the most common drivers are update dependencies (GHA and associated caching parameters) and improve cache , followed by bot alert, extend cache, and fix caching issues. Lower-frequency drivers include migration, clean up, and remove tests. This indicates that cache modifications are not random edits. Instead, they are largely responses to recurring maintenance and workflow needs.

\vspace{0.2cm}

\noindent \textbf{Observation 3.2: \textit{Automation explains many cache version upgrades, while most other cache modifications are driven by human workflow needs.}} Table~\ref{tab:caching_taxonomy} shows that bot-initiated changes appear primarily under bot alert category. Table~\ref{tab:my-table} shows that all bot alert cases map to \textit{cache version update} (40 of 40 cases, 100\%), and bot alerts do not appear as a driver for other targets in the coded sample. In contrast, Table~\ref{tab:caching_taxonomy} shows that \textit{improve cache}, \textit{fix caching issues}, \textit{extend cache}, \textit{migration}, and \textit{clean up} are human-driven, and Table~\ref{tab:my-table} shows these drivers distribute across parameter changes and add/remove actions. This indicates that bot activity is concentrated on version updates, whereas the broader evolution of caching configurations—parameter tuning, adding/removing caches, and restructuring workflows occurs primarily because developers respond to workflow behavior, CI needs, and caching stability issues.

\vspace{0.2cm}

\noindent \textbf{Observation 3.3: \textit{Parameter updates are driven by humans.}} Table~\ref{tab:rq3} shows that transitions targeting \textit{parameter update} (P\_up) dominate the transition data. The most frequent repeated behavior includes P\_up to P\_up (Table~\ref{tab:rq3}). Table~\ref{tab:my-table} shows that when the stated driver is \textit{fix caching issues}, the target is most commonly \textit{parameter update} (16 cases, 69.6\%). Table~\ref{tab:rq3_table_1} further shows that these P\_up cases are human-initiated and are frequently explained by \textit{update dependencies} and \textit{fix caching issues}. This explains the rapid iteration patterns observed in RQ2 (especially in build and test jobs), many frequent parameter updates reflect developers repeatedly adjusting keys/paths and related parameters to stabilize caching behavior and address cache misses or workflow failures, rather than only planned performance refinement.

\vspace{0.2cm}

\noindent \textbf{Observation 3.4: \textit{Cache version updates occur after longer gaps and changes are driven by bot alerts and human driven dependency maintenance of GHA.}} Along with fast parameter updates, RQ2 showed that initial version updates (EC $\rightarrow$ C\_up) are long-delayed actions. Table \ref{tab:rq3} confirms this, showing the median time from enable cache to the first cache version update is 384.13 days. Table \ref{tab:rq3_table_1} explains this delay as the cache version update action (N=86) is almost exclusively an \textbf{Upgrade} (N=84). The drivers for this upgrade are split between human-led updating dependency and automated bot alert. These evidences indicate that developers tend to stay in their initial cache versions. Also, while bots ensure dependency currency, developers also depend on manual PRs to update caching versions. Although rare, downgrade cases also appear, showing that some upgrades introduce problems and later require recovery.

\vspace{0.2cm}

\noindent \textbf{Observation 3.5: \textit{Developers add additional cache to expand and improve their existing workflow.}} Table~\ref{tab:rq3_table_1} shows that for \textit{adding cache}, the most common drivers are \textit{Extend cache} and \textit{Improve cache}, followed by smaller contributions from \textit{fix caching issues}, \textit{fix security issues}, and \textit{migration}. These driver distributions are consistent with the RQ2 observation that $EC \rightarrow C\_add$ occurs within weeks in build and test jobs, suggesting incremental adoption. Developers often add caches to expand CI coverage and improve performance beyond the initial configuration. This supports the view that caching usage evolves through gradual refinement after initial enablement.

\vspace{0.2cm}

\noindent \textbf{Observation 3.6: \textit{Cache removal appears both as a short-gap reversal after cache addition and as a later maintenance action.}} Table~\ref{tab:rq3} shows that \textit{adding cache} $\rightarrow$ \textit{removing cache} occurs with a short cadence (median 4.25 days). Table~\ref{tab:rq3} also shows a longer pattern for $EC \rightarrow C\_rm$, with N=111 and a median of 129.68 days. Table~\ref{tab:rq3_table_1} shows that \textit{improve cache} is a common driver for removals that follow additions, while later removals are frequently explained by \textit{Remove tests} (N=13) and \textit{clean up}. Table~\ref{tab:rq3_table_1} also shows contributions from \textit{migration} and explicit \textit{disable cache}. This shows that cache removal is not explained by one reason. Some removals reflect a quick reversal after a cache is added, while others occur months later when workflows change.

\vspace{0.2cm}

\noindent \textbf{Observation 3.7: \textit{Security-driven cache modifications are infrequent but involve multiple types of actions.}} Table~\ref{tab:caching_taxonomy} shows \textit{fix security issues} is relatively rare. Table~\ref{tab:my-table} shows that these cases map to \textit{adding cache} (N=3, 50.0\%), \textit{parameter update} (N=2, 33.3\%), and \textit{cache version update} (N=1, 16.7\%). When security concerns appear in commit/PR context, developers respond through different workflow edits (adding caches, adjusting parameters, or updating versions) rather than relying on a single type of cache modification. 



\vspace{0.2cm}

\begin{tcolorbox}[enhanced,width=4.7in,size=fbox,drop shadow southwest,sharp corners]
\textbf{RQ3 Summary: Parameter updates dominate cache-related modifications and are commonly performed by developers to fix caching issues, leading to repeated edits to cache keys and paths within short gaps. In contrast, cache version updates typically occur long after enabling caching and are mainly explained by updating dependencies, with bot alerts initiating many of these upgrades. Beyond updates, developers add caches mainly to extend cache usage and improve cache behavior, while cache removal reflects two common situations. Some caches are removed soon after being added due to ineffectiveness and others are removed later when workflows change through removing tests, clean up, migration, or disabling caching. Overall, bots are mainly responsible for cache version updates, whereas developers perform most of the remaining cache modifications that shape how caching evolves in practice. }
\end{tcolorbox}

\section{Implications} \label{sec:disc}

\noindent \textbf{Caching is not a one-time optimization and it requires ongoing maintenance.} Caching is often presented as a straightforward way to reduce CI runtime \citep{bouzenia2024resource}, but our results show that in practice it behaves more like a maintained subsystem than a lightweight optimization flag. RQ2 shows that after enablement, caching evolves through repeated updates, additions, and removals across job types (Figures~\ref{fig:build_tran} to \ref{fig:trans_sync}). In particular, repetitive self-loops such as parameter and cache version updates dominate the transition structure (e.g., build and test in Tables~\ref{tab:build_trans_short} and \ref{tab:test_trans_short}), showing that developers do not simply enable caching once and leave it unchanged. RQ3 reinforces this by showing that the dominant target of maintenance is \textit{parameter update} with 854 transitions (58.20\%) in Table~\ref{tab:rq3}. Taken together, these findings imply that caching should not be described merely as a performance optimization. \textbf{Hence, caching should be treated as an evolving CI configuration component that introduces ongoing maintenance work, corrective edits, and long-term upkeep.}

\vspace{0.2cm}

\noindent \textbf{Automation support in the current bot ecosystem is largely limited to cache version updates.} Our qualitative results show that automation plays a visible but fundamentally narrow role in cache maintenance. Table~\ref{tab:caching_taxonomy} shows that \textit{bot alert} accounts for 40 cases, while the remaining drivers are primarily human-centered, such as \textit{update dependencies} (91), \textit{improve cache} (47), \textit{extend cache} (28), and \textit{fix caching issues} (23). Table~\ref{tab:my-table} further shows that bot alerts map entirely to \textit{cache version update} (40/40, 100\%) and do not appear as a driver for parameter updates, adding cache, removing cache, or cleanup-related actions. This indicates that existing bot ecosystem helps repositories stay current with cache-action versions, but does not reduce the dominant human effort involved in tuning, restructuring, and stabilizing cache configurations. \textbf{Hence, tool builders should move beyond version bumping and develop workflow-aware bots that can assist with broader cache maintenance tasks, especially those involving parameter refinement and corrective reconfiguration.}

\vspace{0.2cm}

\noindent \textbf{Parameter updates are the central maintenance bottleneck in caching and require intelligent tool support.} Among all target states, \textit{parameter update} is the largest maintenance category, accounting for 854 transitions (58.20\%) in Table~\ref{tab:rq3}. This is not only frequent, but also highly repetitive. Table~\ref{tab:rq3} shows that $P\_up \rightarrow P\_up$ alone contributes 497 transitions, and RQ2 shows that these repeated updates occur quickly in build and test jobs, with median delays of 4.08 days and 2.93 days, respectively (Tables~\ref{tab:build_trans_short} and \ref{tab:test_trans_short}; Figures~\ref{fig:build_tran} and \ref{fig:test_tran}). RQ3 further shows that these updates are primarily \textbf{human-driven} and commonly tied to \textit{fix caching issues} and \textit{update dependencies} (Tables~\ref{tab:caching_taxonomy}, \ref{tab:rq3_table_1}, and \ref{tab:my-table}). This implies that the main engineering burden of caching lies not in enabling cache, but in repeatedly refining keys, paths, and related configuration details. Therefore, future agentic and recommendation-based support should prioritize parameter synthesis, cache-key validation, restore-key design, because this is where the largest share of human maintenance effort is currently spent.

\vspace{0.2cm}

\noindent \textbf{Effective cache maintenance support must include helping developers remove unnecessary cache configurations.} Our findings show that caching involves not only introducing and updating cache-related settings, but also removing them as workflows evolve. In RQ2, removal-related transitions appear across multiple job types, showing that previously introduced cache-related configurations are sometimes revised through deletion. RQ3 explains that \textit{Removing cache} is commonly driven by \textit{Improve cache}, \textit{Remove tests}, \textit{Clean up}, \textit{Migration}, and \textit{Disable cache}, whereas \textit{Removing parameter} is associated with \textit{Migration}, \textit{Update dependencies}, \textit{Clean up}, and \textit{Improve cache} (Table~\ref{tab:rq3_table_1}). Table~\ref{tab:my-table} further confirms that these removal-related changes arise from multiple human-driven maintenance needs, rather than from a single recurring trigger. Therefore, future caching tools should not assume that more cache-related configuration is always beneficial. 


\vspace{0.2cm}


\noindent \textbf{Future work should quantify whether frequent cache maintenance actually improves CI execution time.} Our findings show that developers engage in highly frequent cache maintenance, particularly repeated parameter updates in build and test jobs that occur every few days (Tables~\ref{tab:build_trans_short} and \ref{tab:test_trans_short}). However, while our study reveals how often these configurations are tuned, it remains unclear whether such repeated adjustments translate into actual performance gains. As the primary goal of caching is to accelerate CI pipelines, future work should evaluate the resulting change in workflow execution time after specific cache-related transitions. By linking maintenance actions to before-and-after workflow run durations, researchers can determine which configuration changes yield measurable performance improvements and which primarily introduce maintenance overhead without corresponding execution-time benefits.

\section{Threats to Validity} \label{sec:threats}

\noindent \textbf{Construct validity:} Our operationalization of ``caching adoption'' and cache maintenance activities may not capture all ways developers implement caching in GHA workflows. We identify caching usage through cache-related configuration steps and classify step-level strategies into explicit caching, package manager caching, and Docker layer caching. This approach may miss cases where caching is implemented through custom scripts or actions that do not match our detection rules, and it may also include borderline cases that resemble caching but are used for other purposes. Similarly, our job type categorization relies on developer-defined job names and a keyword-based taxonomy. Because job names are inconsistent across repositories, some jobs may be misclassified, which can affect job-level prevalence results in RQ1 and job-type comparisons in RQ2. 

\vspace{0.2cm}

\noindent \textbf{Internal validity:}  We prune low probability transitions below 0.05 to focus on recurring patterns. This design choice may bias the representation of evolutionary patterns by removing rare yet meaningful behaviors. Open coding was the primary methodology employed in our study. This approach introduces subjectivity into our results, as annotators may have different interpretations of the coding scheme. This could introduce annotator bias into our qualitative findings, particularly in the distribution of drivers for cache modifications.  In RQ3, we infer drivers from commit messages and, when available, linked PR context. When the rationale is missing or unclear, the transition is coded as Other and excluded from driver-focused interpretation. This introduces a threat that some drivers are underrepresented because they are not explicitly documented in the commit or PR text. 
\vspace{0.2cm}

\noindent \textbf{Conclusion validity:} Our findings rely primarily on descriptive statistics, transition probabilities, and time to transition summaries. Some job types have fewer observations, which can increase variance in estimated probabilities and medians and can make long tail behaviors sensitive to small changes in the dataset. For the qualitative component, we use a statistically significant random sample per target state with a 90\% confidence interval and 10\% margin of error. While this sampling supports stable estimates for common drivers, rare drivers and rare actions such as downgrades may remain sensitive to sampling noise. We mitigate over interpretation by reporting exact counts and by treating low frequency categories as limited evidence rather than broad trends.

\vspace{0.2cm}

\noindent \textbf{External validity:} Our dataset consists of GitHub repositories and workflows collected using the GitHub GraphQL API and reconstructed workflow histories, as described in Section~\ref{sec:dataset}. The observed adoption rates, job type prevalence, and maintenance patterns may not generalize to private repositories, enterprise environments, or repositories that use different CI platforms. Results may also differ across language ecosystems, workflow templates, and project domains that are not represented in our sample. In addition, our step level strategy taxonomy reflects the caching mechanisms observed in our dataset. Other caching approaches may exist and could be more common in different populations. Therefore, we interpret our results as evidence about caching adoption and evolution within the studied GHA sample rather than as universal properties of CI caching.

\vspace{-0.4cm}
\section{Related Work}\label{sec:related}
\subsection{Studies on GitHub Actions}

GitHub Actions (GHA), introduced in 2019, has rapidly become a major automation platform in open-source software development.\footnote{https://github.blog/news-insights/product-news/github-actions-now-support-ci-cd/} Early empirical studies established its widespread adoption. For example, Decan et al. analyzed 68K repositories and reported that 43.9\% used GHA, while also identifying common workflow triggers and automation practices \citep{decan2022use}. Similarly, Chen et al. found broad adoption of GHA across the projects they studied \citep{chen2021let}. Together, these studies show that GHA has become a core part of continuous integration and deployment (CI/CD) processes.

As adoption increased, researchers began to investigate how GHA workflows evolve over time. Valenzuela et al. manually analyzed 222 commits and identified 11 categories of workflow modifications, showing that workflow maintenance is frequent and multifaceted \citep{valenzuela2022evolution}. In subsequent work, they highlighted the hidden costs of workflow maintenance, showing that bug fixing and CI/CD performance improvement are major drivers of workflow changes \citep{valenzuela2024hidden}. Rostami Mazrae et al. conducted a mixed-method study of workflow evolution using 439 manually analyzed modified workflow files and a large-scale quantitative analysis over 49K+ repositories \citep{rostami5369484empirical}. Their results show that GHA workflows are continuously maintained, with modifications dominating over additions and removals, and that most changes concentrate on task specification and task configuration, especially within workflow jobs and steps. Complementing this line of work, Zheng et al. conducted a large-scale empirical study of workflow failures and identified recurring failure categories and configuration weaknesses that reduce workflow reliability \citep{zheng2025github}. Huang et al. studied rerun practices in GitHub Actions workflows, focusing on cases where developers rerun workflows or failed jobs without modifying the repository \citep{huang2026reruns}. Their study of 3,320 open-source Java repositories quantified the time and computing waste associated with reruns, analyzed cases where reruns succeed, and investigated root causes of workflow execution flakiness. Their results provide insight into how developers cope with workflow failures and highlight reruns as an important reliability and efficiency concern in GitHub Actions.

Other studies have focused on the structural, operational, and security characteristics of GHA workflows. Abrokwah et al. examined workflow heterogeneity, compliance, and complexity, showing substantial structural variation and deviation from best practices across repositories \citep{abrokwah2025empirical}. Rostami et al. classified workflow changes at scale, revealing systematic patterns of addition, refactoring, and removal over time \citep{rostami5369484empirical}. Bouzenia et al. quantified resource usage in GHA and identified opportunities to reduce workflow cost and execution time \citep{bouzenia2024resource}. From a security perspective, prior work has shown that many workflows depend on outdated third-party Actions, exposing repositories to known vulnerabilities \citep{decan2023outdatedness, benedetti2022automatic}. Decan et al. further showed that the reuse of Actions is highly concentrated on a small subset of the ecosystem \citep{decan2023outdatedness}. Beyond workflow internals, Wessel et al. demonstrated that adopting GHA can alter pull request dynamics, including review time and communication behavior, and also examined the broader automation and bot ecosystem enabled by GHA \citep{wessel2023github}.

\vspace{0.1cm}
In summary, existing research has studied GHA from multiple perspectives, including adoption, maintenance, failures, reruns, complexity, performance, and security. However, caching has not been examined as a distinct topic of empirical study. When discussed, it is usually presented as only one example of workflow optimization and maintenance. In contrast, our work provides a cache-centric and fine-grained view of GHA workflows. Specifically, we examine caching at the \textit{job} and \textit{step} levels, analyze where cache-related configurations appear across different CI/CD phases, and study how these configurations evolve over time through concrete cache-related activities. We further investigate the motivations behind these changes and identify who initiates them. This makes our work more granular than prior GHA studies on workflow evolution and also distinct from prior CI caching studies that focus mainly on overall build-level adoption and performance outcomes.

\subsection{Studies on CI/CD Caching Strategies}

Caching is a widely used acceleration technique in software build and CI/CD pipelines, but its effectiveness depends on what is cached, how the cache is maintained, and where it is applied in the pipeline. Prior work has shown that caching can substantially reduce build time \citep{bouzenia2024resource}, \citep{ghaleb2026promise}, while also introducing trade-offs related to invalidation, download overhead, and maintenance complexity\footnote{https://www.datadoghq.com/blog/cache-purge-ci-cd/}. Gallaba et al. empirically studied environment caching and step-skipping, showing that dependency and environment caching can accelerate builds, but that the benefits may diminish when cache retrieval overhead or invalidation costs are high \citep{gallaba2020accelerating}. In the Bazel ecosystem, Zheng et al. conducted controlled experiments across hundreds of projects, evaluating parallel and incremental builds under popular CI services, including GitHub Actions, and found that remote and incremental caching can improve performance, although the gains are highly workload-dependent \citep{zheng2024does}. Ghaleb et al. studied CI caching in Travis CI and showed that caching adoption, maintenance, and performance benefits vary substantially across projects, with many projects applying caching only once and seeing limited or inconsistent build-time improvements \citep{ghaleb2026promise}.  These studies provide valuable evidence that caching can improve CI efficiency, but they primarily focus on performance outcomes, cache overhead under specific tools, build systems, and experimental settings.

In summary, prior work on caching has mainly evaluated whether particular caching mechanisms improve performance, when they are beneficial, and what overheads they introduce. Our work complements this literature by studying caching as it is configured and maintained \emph{in the wild} within GHA workflows. Unlike prior work on Travis CI caching, which primarily studies caching at the project and build levels, our work investigates caching in GHA at finer granularity, focusing on job- and step-level configurations. This allows us to examine where caching is introduced across workflow phases, how cache-related configurations evolve through concrete maintenance operations over time, and why and by whom these changes are made. Thus, rather than focusing only on whether caching improves runtime, we study caching as an evolving workflow configuration practice in modern GHA workflows.

\section{Conclusion and Future Work}\label{sec:conclusion}

This paper presented a comprehensive characterization of caching adoption and evolution in GitHub Actions workflows. We found that repositories adopting caching are typically larger and more active than non-adopters, often managing their CI workloads with a single workflow file. Within these repositories, caching is most prevalent in build and test jobs where developers predominantly rely on explicit caching strategies over package manager caching. Our evolutionary analysis reveals that caching is not a one-time configuration, it requires repetitive maintenance activities. We observed that build and test jobs undergo frequent and rapid parameter tuning as developers iteratively adjust cache keys and paths to address caching issues. in contrast, integration and release jobs remain stable for longer periods. Through our qualitative analysis of drivers, we determined that while bots play a significant role in triggering long-delayed cache version updates, humans are responsible for the majority of maintenance activities. These human-driven modifications focus on extending cache coverage, improving cache performance, and fixing caching issues through repeated parameter updates.

In the future, we plan to expand our work by conducting an impact-driven analysis that correlates the evolutionary events we identified with build log data, allowing us to quantify the direct effect of these changes on build duration, success rates, and resource costs. Furthermore, given our finding that humans handle all complex tasks, there is a clear need for better automated support. We therefore plan to design and develop specialized tools for cache maintenance to reduce repetitive trial-and-error edits and help developers keep caching effective over time.

\begin{acknowledgements}
We thank Vu Thanh Loc Mai and Mohammad Sadegh Sheikhaei for their help in this project. Also, we acknowledge the support of the Natural Sciences and Engineering Research Council of Canada (NSERC), [funding reference number: RGPIN-2025-04723]. 

\end{acknowledgements}

\section*{Conflict of Interest}
The authors declare that they have no conflict of interest.



\bibliographystyle{spbasic}      

\bibliography{references}
\appendix
\section{Evolution of caching configurations for different job type}\label{sec:appen}
\begin{table}[h]
\centering
\caption{Evolution of caching configurations for the build job type. For each source state (From), the table details the resulting target states (To), the total occurrences, and the conditional probability (\%) of each specific transition. The remaining columns report the descriptive statistics (Avg, Med, Std, Min, Max) for the time in days elapsed before each transition.}
\label{tab:build_trans_short}
\resizebox{\linewidth}{!}{%
\begin{tabular}{llrrrrrrrr}
\toprule
\textbf{From} & \textbf{To} & \textbf{Count} & \textbf{Total} & \textbf{Avg} & \textbf{Med} & \textbf{Std} & \textbf{Min} & \textbf{Max} & \textbf{\%} \\ 
\midrule
C\_add & C\_add & 26 & \multirow{4}{*}{75} & 28.03 & 0.00 & 93.91 & 0.00 & 462.91 & 34.67 \\ 
 & C\_rm & 19 & & 112.31 & 0.00 & 219.41 & 0.00 & 750.21 & 25.33 \\ 
 & P\_up & 16 & & 106.70 & 51.16 & 132.88 & 0.05 & 419.07 & 21.33 \\ 
 & C\_up & 14 & & 373.24 & 309.69 & 342.69 & 0.00 & 1080.84 & 18.67 \\ 
\midrule
P\_add & C\_up & 3 & \multirow{4}{*}{7} & 59.94 & 0.00 & 103.82 & 0.00 & 179.82 & 42.86 \\ 
 & P\_up & 2 & & 3.16 & 3.16 & 4.46 & 0.00 & 6.31 & 28.57 \\ 
 & P\_add & 1 & & 0.00 & 0.00 & -- & 0.00 & 0.00 & 14.29 \\ 
 & P\_rm & 1 & & 0.00 & 0.00 & -- & 0.00 & 0.00 & 14.29 \\ 
\midrule
C\_up & C\_up & 177 & \multirow{3}{*}{222} & 59.56 & 0.00 & 138.32 & 0.00 & 668.08 & 77.63 \\ 
 & P\_up & 27 & & 148.09 & 79.07 & 179.24 & 0.00 & 592.69 & 11.84 \\ 
 & C\_rm & 18 & & 270.58 & 328.70 & 225.37 & 0.00 & 707.67 & 7.89 \\ 
\midrule
EC & C\_up & 69 & \multirow{5}{*}{231} & 637.14 & 587.98 & 448.74 & 6.35 & 1615.93 & 29.49 \\ 
 & EC & 62 & & 0.00 & 0.00 & 0.00 & 0.00 & 0.00 & 26.50 \\ 
 & P\_up & 40 & & 173.07 & 105.93 & 221.14 & 0.00 & 874.04 & 17.09 \\ 
 & C\_rm & 39 & & 422.58 & 343.82 & 428.02 & 0.00 & 1514.81 & 16.67 \\ 
 & C\_add & 21 & & 153.44 & 53.14 & 177.26 & 0.00 & 504.87 & 8.97 \\ 
\midrule
C\_rm & C\_rm & 48 & \multirow{4}{*}{83} & 13.73 & 0.00 & 66.32 & 0.00 & 390.89 & 55.17 \\ 
 & C\_add & 19 & & 47.11 & 0.01 & 105.93 & 0.00 & 445.13 & 21.84 \\ 
 & P\_up & 9 & & 55.79 & 30.82 & 101.96 & 0.00 & 321.98 & 10.34 \\ 
 & C\_up & 7 & & 220.73 & 0.00 & 510.42 & 0.00 & 1374.41 & 8.05 \\ 
\midrule
P\_rm & C\_rm & 2 & \multirow{2}{*}{4} & 0.00 & 0.00 & 0.00 & 0.00 & 0.00 & 50.00 \\ 
 & P\_up & 2 & & 117.53 & 117.53 & 166.21 & 0.00 & 235.06 & 50.00 \\ 
\midrule
P\_up & P\_up & 135 & \multirow{4}{*}{201} & 45.40 & 4.08 & 105.61 & 0.00 & 941.90 & 65.53 \\ 
 & C\_up & 31 & & 173.03 & 104.80 & 277.52 & 0.00 & 1366.00 & 15.05 \\ 
 & C\_rm & 24 & & 152.99 & 43.29 & 222.58 & 0.00 & 902.66 & 11.65 \\ 
 & C\_add & 11 & & 25.17 & 4.73 & 52.00 & 1.74 & 177.96 & 5.34 \\ 
\bottomrule
\end{tabular}%
}
\end{table}


\begin{table}[h]
\centering
\caption{Evolution of caching configurations for the test job type. For each source state (From), the table details the resulting target states (To), the total occurrences, and the conditional probability (\%) of each specific transition. The remaining columns report the descriptive statistics (Avg, Med, Std, Min, Max) for the time in days elapsed before each transition.}
\label{tab:test_trans_short}
\resizebox{\linewidth}{!}{%
\begin{tabular}{llrrrrrrrr}
\toprule
\textbf{From} & \textbf{To} & \textbf{Count} & \textbf{Total} & \textbf{Avg} & \textbf{Med} & \textbf{Std} & \textbf{Min} & \textbf{Max} & \textbf{\%} \\ 
\midrule
P\_up & P\_up & 142 & \multirow{4}{*}{208} & 62.75 & 2.93 & 156.46 & 0.00 & 986.95 & 66.05 \\ 
 & C\_up & 31 & & 111.24 & 31.50 & 152.83 & 0.00 & 584.03 & 14.42 \\ 
 & C\_rm & 19 & & 221.16 & 7.24 & 393.51 & 0.00 & 1154.81 & 8.84 \\ 
 & C\_add & 16 & & 64.67 & 43.66 & 85.87 & 0.00 & 323.27 & 7.44 \\ 
\midrule
C\_up & C\_up & 114 & \multirow{3}{*}{159} & 58.67 & 0.00 & 137.94 & 0.00 & 713.59 & 68.26 \\ 
 & P\_up & 33 & & 76.46 & 0.00 & 120.17 & 0.00 & 433.71 & 19.76 \\ 
 & C\_rm & 12 & & 150.41 & 94.56 & 148.67 & 0.00 & 435.59 & 7.19 \\ 
\midrule
C\_add & C\_add & 48 & \multirow{5}{*}{133} & 24.83 & 0.00 & 93.51 & 0.00 & 463.22 & 36.09 \\ 
 & C\_rm & 42 & & 112.53 & 32.49 & 152.93 & 0.00 & 485.66 & 31.58 \\ 
 & P\_up & 21 & & 105.59 & 8.47 & 219.12 & 0.00 & 973.79 & 15.79 \\ 
 & C\_up & 15 & & 340.95 & 231.83 & 356.57 & 0.01 & 1294.85 & 11.28 \\ 
 & P\_add & 7 & & 0.00 & 0.00 & 0.00 & 0.00 & 0.00 & 5.26 \\ 
\midrule
EC & P\_up & 45 & \multirow{5}{*}{183} & 263.33 & 85.03 & 347.74 & 0.00 & 981.44 & 24.32 \\ 
 & C\_up & 41 & & 515.49 & 351.05 & 457.35 & 0.20 & 1549.93 & 22.16 \\ 
 & EC & 38 & & 0.00 & 0.00 & 0.00 & 0.00 & 0.00 & 20.54 \\ 
 & C\_rm & 30 & & 243.13 & 120.77 & 304.07 & 0.00 & 1271.64 & 16.22 \\ 
 & C\_add & 29 & & 172.12 & 41.97 & 243.45 & 0.01 & 962.46 & 15.68 \\ 
\midrule
C\_rm & C\_rm & 44 & \multirow{5}{*}{103} & 39.12 & 0.00 & 105.53 & 0.00 & 448.01 & 41.51 \\ 
 & C\_add & 32 & & 21.35 & 0.00 & 80.79 & 0.00 & 445.13 & 30.19 \\ 
 & P\_rm & 10 & & 0.00 & 0.00 & 0.00 & 0.00 & 0.00 & 9.43 \\ 
 & P\_up & 9 & & 85.85 & 13.21 & 189.00 & 0.00 & 584.88 & 8.49 \\ 
 & C\_up & 8 & & 265.57 & 269.97 & 180.03 & 0.00 & 557.51 & 7.55 \\ 
\midrule
P\_rm & C\_add & 8 & \multirow{3}{*}{13} & 0.00 & 0.00 & 0.00 & 0.00 & 0.00 & 61.54 \\ 
 & C\_rm & 3 & & 0.00 & 0.00 & 0.00 & 0.00 & 0.00 & 23.08 \\ 
 & C\_up & 2 & & 0.00 & 0.00 & 0.00 & 0.00 & 0.00 & 15.38 \\ 
\midrule
P\_add & C\_add & 5 & \multirow{6}{*}{18} & 0.00 & 0.00 & 0.00 & 0.00 & 0.00 & 27.78 \\ 
 & P\_up & 5 & & 186.87 & 140.78 & 257.22 & 0.00 & 628.32 & 27.78 \\ 
 & P\_add & 3 & & 0.00 & 0.00 & 0.00 & 0.00 & 0.00 & 16.67 \\ 
 & C\_up & 3 & & 0.00 & 0.00 & 0.00 & 0.00 & 0.00 & 16.67 \\ 
 & C\_rm & 1 & & 875.94 & 875.94 & - & 875.94 & 875.94 & 5.56 \\ 
 & EC & 1 & & 0.00 & 0.00 & - & 0.00 & 0.00 & 5.56 \\ 
\bottomrule
\end{tabular}%
}
\end{table}


\begin{table}[h]
\centering
\caption{Evolution of caching configurations for the integration job type. For each source state (From), the table details the resulting target states (To), the total occurrences, and the conditional probability (\%) of each specific transition. The remaining columns report the descriptive statistics (Avg, Med, Std, Min, Max) for the time in days elapsed before each transition.}
\label{tab:integration_trans_short}
\resizebox{\linewidth}{!}{%
\begin{tabular}{llrrrrrrrr}
\toprule
\textbf{From} & \textbf{To} & \textbf{Count} & \textbf{Total} & \textbf{Avg} & \textbf{Med} & \textbf{Std} & \textbf{Min} & \textbf{Max} & \textbf{\%} \\ 
\midrule
P\_up & P\_up & 25 & \multirow{2}{*}{35} & 137.45 & 87.25 & 124.44 & 13.03 & 538.19 & 67.57 \\ 
 & C\_up & 10 & & 235.11 & 141.92 & 405.25 & 0.00 & 1366.00 & 27.03 \\ 
\midrule
C\_up & P\_up & 9 & \multirow{2}{*}{10} & 78.27 & 64.22 & 97.34 & 0.00 & 292.84 & 90.00 \\ 
 & C\_rm & 1 & & 285.92 & 285.92 & 0.00 & 285.92 & 285.92 & 10.00 \\ 
\midrule
EC & P\_up & 8 & \multirow{4}{*}{16} & 180.66 & 203.39 & 91.09 & 0.00 & 300.88 & 50.00 \\ 
 & C\_up & 3 & & 474.73 & 326.13 & 304.42 & 273.13 & 824.91 & 18.75 \\ 
 & C\_rm & 3 & & 184.33 & 252.89 & 119.54 & 46.30 & 253.81 & 18.75 \\ 
 & C\_add & 2 & & 438.48 & 438.48 & 250.70 & 261.21 & 615.76 & 12.50 \\ 
\midrule
C\_rm & C\_add & 4 & \multirow{2}{*}{5} & 19.20 & 0.00 & 38.39 & 0.00 & 76.79 & 80.00 \\ 
 & C\_rm & 1 & & 0.00 & 0.00 & 0.00 & 0.00 & 0.00 & 20.00 \\ 
\midrule
C\_add & C\_rm & 4 & \multirow{4}{*}{8} & 82.21 & 1.54 & 162.37 & 0.00 & 325.75 & 50.00 \\ 
 & C\_add & 2 & & 5.21 & 5.21 & 7.37 & 0.00 & 10.42 & 25.00 \\ 
 & C\_up & 1 & & 584.03 & 584.03 & 0.00 & 584.03 & 584.03 & 12.50 \\ 
 & P\_up & 1 & & 8.47 & 8.47 & 0.00 & 8.47 & 8.47 & 12.50 \\ 
\midrule
P\_rm & C\_rm & 1 & 1 & 0.00 & 0.00 & 0.00 & 0.00 & 0.00 & 100.00 \\ 
\midrule
P\_add & C\_up & 1 & \multirow{2}{*}{2} & 584.03 & 584.03 & 0.00 & 584.03 & 584.03 & 50.00 \\ 
 & P\_up & 1 & & 161.81 & 161.81 & 0.00 & 161.81 & 161.81 & 50.00 \\ 
\bottomrule
\end{tabular}%
}
\end{table}


\begin{table}[h]
\centering
\caption{Evolution of caching configurations for the release job type. For each source state (From), the table details the resulting target states (To), the total occurrences, and the conditional probability (\%) of each specific transition. The remaining columns report the descriptive statistics (Avg, Med, Std, Min, Max) for the time in days elapsed before each transition.}
\label{tab:release_trans_short}
\resizebox{\linewidth}{!}{%
\begin{tabular}{llrrrrrrrr}
\toprule
\textbf{From} & \textbf{To} & \textbf{Count} & \textbf{Total} & \textbf{Avg} & \textbf{Med} & \textbf{Std} & \textbf{Min} & \textbf{Max} & \textbf{\%} \\ 
\midrule
P\_up & P\_up & 30 & \multirow{3}{*}{45} & 90.86 & 61.95 & 86.46 & 0.00 & 311.97 & 63.83 \\ 
 & C\_up & 11 & & 166.19 & 137.67 & 153.70 & 0.00 & 557.50 & 23.40 \\ 
 & C\_rm & 4 & & 91.89 & 85.88 & 31.95 & 59.81 & 135.99 & 8.51 \\ 
\midrule
EC & C\_rm & 18 & \multirow{5}{*}{53} & 273.29 & 85.34 & 377.79 & 0.00 & 1170.03 & 33.96 \\ 
 & C\_up & 13 & & 445.71 & 270.37 & 473.25 & 0.25 & 1298.49 & 24.53 \\ 
 & P\_up & 12 & & 203.64 & 125.98 & 197.30 & 0.00 & 512.00 & 22.64 \\ 
 & C\_add & 5 & & 71.33 & 11.93 & 133.69 & 1.08 & 309.99 & 9.43 \\ 
 & EC & 5 & & 0.00 & 0.00 & 0.00 & 0.00 & 0.00 & 9.43 \\ 
\midrule
C\_up & C\_up & 13 & \multirow{3}{*}{25} & 137.12 & 0.00 & 200.97 & 0.00 & 520.81 & 50.00 \\ 
 & P\_up & 8 & & 70.56 & 18.06 & 107.20 & 0.00 & 292.02 & 30.77 \\ 
 & C\_rm & 4 & & 371.48 & 429.10 & 238.07 & 34.54 & 593.18 & 15.38 \\ 
\midrule
C\_rm & C\_rm & 10 & \multirow{4}{*}{18} & 14.50 & 0.00 & 45.86 & 0.00 & 145.03 & 55.56 \\ 
 & C\_add & 6 & & 74.19 & 0.00 & 181.73 & 0.00 & 445.13 & 33.33 \\ 
 & C\_up & 1 & & 35.09 & 35.09 & 0.00 & 35.09 & 35.09 & 5.56 \\ 
 & P\_up & 1 & & 36.92 & 36.92 & 0.00 & 36.92 & 36.92 & 5.56 \\ 
\midrule
C\_add & C\_add & 4 & \multirow{5}{*}{12} & 0.00 & 0.00 & 0.00 & 0.00 & 0.00 & 33.33 \\ 
 & C\_up & 4 & & 150.73 & 117.86 & 172.97 & 0.04 & 367.15 & 33.33 \\ 
 & C\_rm & 2 & & 0.49 & 0.49 & 0.70 & 0.00 & 0.99 & 16.67 \\ 
 & P\_add & 1 & & 0.42 & 0.42 & 0.00 & 0.42 & 0.42 & 8.33 \\ 
 & P\_up & 1 & & 55.04 & 55.04 & 0.00 & 55.04 & 55.04 & 8.33 \\ 
\midrule
P\_add & C\_up & 1 & 1 & 0.00 & 0.00 & 0.00 & 0.00 & 0.00 & 100.00 \\ 
\bottomrule
\end{tabular}%
}
\end{table}


\begin{table}[h]
\centering
\caption{Evolution of caching configurations for the lint job type. For each source state (From), the table details the resulting target states (To), the total occurrences, and the conditional probability (\%) of each specific transition. The remaining columns report the descriptive statistics (Avg, Med, Std, Min, Max) for the time in days elapsed before each transition.}

\label{tab:lint_trans_short}
\resizebox{\linewidth}{!}{%
\begin{tabular}{llrrrrrrrr}
\toprule
\textbf{From} & \textbf{To} & \textbf{Count} & \textbf{Total} & \textbf{Avg} & \textbf{Med} & \textbf{Std} & \textbf{Min} & \textbf{Max} & \textbf{\%} \\ 
\midrule
EC & C\_up & 11 & \multirow{4}{*}{21} & 476.04 & 160.59 & 530.04 & 0.01 & 1413.24 & 47.83 \\ 
 & C\_rm & 4 & & 428.71 & 340.82 & 382.73 & 115.27 & 917.91 & 17.39 \\ 
 & P\_up & 4 & & 263.36 & 99.76 & 380.50 & 23.89 & 830.04 & 17.39 \\ 
 & EC & 2 & & 0.00 & 0.00 & 0.00 & 0.00 & 0.00 & 8.70 \\ 
\midrule
P\_up & C\_up & 8 & \multirow{3}{*}{18} & 244.19 & 134.07 & 247.46 & 11.38 & 648.94 & 44.44 \\ 
 & P\_up & 8 & & 116.10 & 103.77 & 83.99 & 27.75 & 263.80 & 44.44 \\ 
 & C\_rm & 2 & & 535.55 & 535.55 & 470.44 & 202.90 & 868.20 & 11.11 \\ 
\midrule
C\_up & C\_up & 8 & \multirow{4}{*}{19} & 188.50 & 190.79 & 185.58 & 0.00 & 508.01 & 38.10 \\ 
 & P\_up & 7 & & 60.67 & 0.03 & 88.67 & 0.00 & 224.30 & 33.33 \\ 
 & C\_rm & 2 & & 2.04 & 2.04 & 2.88 & 0.00 & 4.08 & 9.52 \\ 
 & P\_rm & 2 & & 0.00 & 0.00 & 0.00 & 0.00 & 0.00 & 9.52 \\ 
\midrule
C\_rm & C\_rm & 7 & \multirow{3}{*}{11} & 0.00 & 0.00 & 0.00 & 0.00 & 0.00 & 58.33 \\ 
 & C\_add & 2 & & 222.57 & 222.57 & 314.76 & 0.00 & 445.13 & 16.67 \\ 
 & P\_rm & 2 & & 0.00 & 0.00 & 0.00 & 0.00 & 0.00 & 16.67 \\ 
\midrule
C\_add & C\_add & 3 & \multirow{4}{*}{7} & 0.00 & 0.00 & 0.00 & 0.00 & 0.00 & 42.86 \\ 
 & C\_up & 2 & & 201.12 & 201.12 & 234.80 & 35.09 & 367.15 & 28.57 \\ 
 & P\_add & 1 & & 0.00 & 0.00 & 0.00 & 0.00 & 0.00 & 14.29 \\ 
 & C\_rm & 1 & & 0.00 & 0.00 & 0.00 & 0.00 & 0.00 & 14.29 \\ 
\midrule
P\_add & C\_up & 2 & \multirow{3}{*}{4} & 814.58 & 814.58 & 779.83 & 263.15 & 1366.00 & 50.00 \\ 
 & C\_add & 1 & & 89.77 & 89.77 & 0.00 & 89.77 & 89.77 & 25.00 \\ 
 & P\_up & 1 & & 0.15 & 0.15 & 0.00 & 0.15 & 0.15 & 25.00 \\ 
\midrule
P\_rm & C\_up & 2 & \multirow{2}{*}{4} & 0.00 & 0.00 & 0.00 & 0.00 & 0.00 & 50.00 \\ 
 & C\_rm & 2 & & 0.00 & 0.00 & 0.00 & 0.00 & 0.00 & 50.00 \\ 
\bottomrule
\end{tabular}%
}
\end{table}


\begin{table}[h]
\centering
\caption{Evolution of caching configurations for the lint job type. For each source state (From), the table details the resulting target states (To), the total occurrences, and the conditional probability (\%) of each specific transition. The remaining columns report the descriptive statistics (Avg, Med, Std, Min, Max) for the time in days elapsed before each transition.}

\label{tab:analyze_trans_short}
\resizebox{\linewidth}{!}{%
\begin{tabular}{llrrrrrrrr}
\toprule
\textbf{From} & \textbf{To} & \textbf{Count} & \textbf{Total} & \textbf{Avg} & \textbf{Med} & \textbf{Std} & \textbf{Min} & \textbf{Max} & \textbf{\%} \\ 
\midrule
P\_up & P\_up & 46 & \multirow{2}{*}{49} & 120.01 & 26.14 & 173.33 & 0.20 & 611.55 & 92.00 \\ 
 & C\_up & 3 & & 269.95 & 257.62 & 276.33 & 0.00 & 552.24 & 6.00 \\ 
\midrule
EC & P\_up & 19 & \multirow{4}{*}{34} & 151.56 & 94.53 & 133.79 & 0.01 & 369.06 & 54.29 \\ 
 & C\_up & 10 & & 517.14 & 384.02 & 542.26 & 0.01 & 1549.93 & 28.57 \\ 
 & C\_rm & 3 & & 196.91 & 3.48 & 338.04 & 0.01 & 587.24 & 8.57 \\ 
 & EC & 2 & & 0.00 & 0.00 & 0.00 & 0.00 & 0.00 & 5.71 \\ 
\midrule
C\_up & C\_up & 6 & \multirow{4}{*}{10} & 199.45 & 190.42 & 218.27 & 0.00 & 431.84 & 60.00 \\ 
 & C\_add & 2 & & 7.10 & 7.10 & 10.05 & 0.00 & 14.21 & 20.00 \\ 
 & C\_rm & 1 & & 4.08 & 4.08 & 0.00 & 4.08 & 4.08 & 10.00 \\ 
 & P\_up & 1 & & 0.00 & 0.00 & 0.00 & 0.00 & 0.00 & 10.00 \\ 
\midrule
C\_rm & C\_rm & 5 & \multirow{3}{*}{10} & 0.00 & 0.00 & 0.00 & 0.00 & 0.00 & 50.00 \\ 
 & C\_add & 4 & & 56.72 & 3.91 & 108.25 & 0.00 & 219.07 & 40.00 \\ 
 & P\_rm & 1 & & 0.00 & 0.00 & 0.00 & 0.00 & 0.00 & 10.00 \\ 
\midrule
C\_add & C\_rm & 3 & \multirow{3}{*}{6} & 251.84 & 250.30 & 132.99 & 119.62 & 385.59 & 50.00 \\ 
 & C\_up & 2 & & 250.48 & 250.48 & 354.23 & 0.00 & 500.96 & 33.33 \\ 
 & P\_up & 1 & & 217.80 & 217.80 & 0.00 & 217.80 & 217.80 & 16.67 \\ 
\midrule
P\_rm & C\_rm & 1 & 1 & 0.00 & 0.00 & 0.00 & 0.00 & 0.00 & 100.00 \\ 
\bottomrule
\end{tabular}%
}
\end{table}


\begin{table}[h]
\centering
\caption{Evolution of caching configurations for the linux job type. For each source state (From), the table details the resulting target states (To), the total occurrences, and the conditional probability (\%) of each specific transition. The remaining columns report the descriptive statistics (Avg, Med, Std, Min, Max) for the time in days elapsed before each transition.}

\label{tab:linux_trans_short}
\resizebox{\linewidth}{!}{%
\begin{tabular}{llrrrrrrrr}
\toprule
\textbf{From} & \textbf{To} & \textbf{Count} & \textbf{Total} & \textbf{Avg} & \textbf{Med} & \textbf{Std} & \textbf{Min} & \textbf{Max} & \textbf{\%} \\ 
\midrule
C\_up & C\_up & 4 & \multirow{2}{*}{5} & 230.93 & 196.61 & 272.48 & 0.00 & 530.50 & 80.00 \\ 
 & P\_up & 1 & & 210.97 & 210.97 & 0.00 & 210.97 & 210.97 & 20.00 \\ 
\midrule
EC & C\_up & 4 & \multirow{3}{*}{7} & 452.34 & 476.09 & 442.51 & 15.77 & 841.42 & 57.14 \\ 
 & C\_add & 2 & & 71.45 & 71.45 & 99.70 & 0.96 & 141.95 & 28.57 \\ 
 & P\_up & 1 & & 133.90 & 133.90 & 0.00 & 133.90 & 133.90 & 14.29 \\ 
\midrule
P\_up & P\_up & 3 & \multirow{2}{*}{5} & 191.71 & 200.53 & 76.88 & 110.81 & 263.80 & 60.00 \\ 
 & C\_up & 2 & & 150.85 & 150.85 & 65.13 & 104.80 & 196.90 & 40.00 \\ 
\midrule
C\_add & C\_up & 1 & 1 & 282.87 & 282.87 & 0.00 & 282.87 & 282.87 & 100.00 \\ 
\bottomrule
\end{tabular}%
}
\end{table}


\begin{table}[h]
\centering
\caption{Evolution of caching configurations for the sync job type. For each source state (From), the table details the resulting target states (To), the total occurrences, and the conditional probability (\%) of each specific transition. The remaining columns report the descriptive statistics (Avg, Med, Std, Min, Max) for the time in days elapsed before each transition.}
\label{tab:sync_trans_short}
\resizebox{\linewidth}{!}{%
\begin{tabular}{llrrrrrrrr}
\toprule
\textbf{From} & \textbf{To} & \textbf{Count} & \textbf{Total} & \textbf{Avg} & \textbf{Med} & \textbf{Std} & \textbf{Min} & \textbf{Max} & \textbf{\%} \\ 
\midrule
EC & P\_up & 28 & \multirow{2}{*}{30} & 250.33 & 72.96 & 284.48 & 0.00 & 693.07 & 90.32 \\ 
 & C\_rm & 2 & & 52.97 & 52.97 & 69.26 & 3.99 & 101.94 & 6.45 \\ 
\midrule
P\_up & P\_up & 18 & \multirow{4}{*}{35} & 409.30 & 288.65 & 220.85 & 0.69 & 698.58 & 48.65 \\ 
 & C\_rm & 11 & & 115.62 & 13.15 & 123.56 & 7.07 & 244.66 & 29.73 \\ 
 & P\_rm & 4 & & 23.02 & 7.07 & 31.89 & 7.07 & 70.85 & 10.81 \\ 
 & C\_add & 2 & & 0.00 & 0.00 & 0.00 & 0.00 & 0.00 & 5.41 \\ 
\midrule
C\_add & C\_add & 11 & \multirow{2}{*}{22} & 0.00 & 0.00 & 0.00 & 0.00 & 0.00 & 45.83 \\ 
 & P\_up & 11 & & 108.01 & 8.47 & 209.03 & 7.78 & 685.42 & 45.83 \\ 
\midrule
P\_rm & C\_add & 6 & \multirow{2}{*}{11} & 0.00 & 0.00 & 0.00 & 0.00 & 0.00 & 54.55 \\ 
 & C\_rm & 5 & & 0.00 & 0.00 & 0.00 & 0.00 & 0.00 & 45.45 \\ 
\midrule
C\_rm & P\_rm & 6 & \multirow{4}{*}{13} & 0.00 & 0.00 & 0.00 & 0.00 & 0.00 & 46.15 \\ 
 & C\_add & 5 & & 0.00 & 0.00 & 0.00 & 0.00 & 0.01 & 38.46 \\ 
 & C\_rm & 1 & & 0.00 & 0.00 & 0.00 & 0.00 & 0.00 & 7.69 \\ 
 & P\_add & 1 & & 0.00 & 0.00 & 0.00 & 0.00 & 0.00 & 7.69 \\ 
\midrule
C\_up & P\_up & 2 & 2 & 0.00 & 0.00 & 0.00 & 0.00 & 0.00 & 100.00 \\ 
\midrule
P\_add & C\_add & 1 & \multirow{2}{*}{2} & 0.00 & 0.00 & 0.00 & 0.00 & 0.00 & 50.00 \\ 
 & C\_up & 1 & & 0.00 & 0.00 & 0.00 & 0.00 & 0.00 & 50.00 \\ 
\bottomrule
\end{tabular}%
}
\end{table}

\end{document}